\documentclass[12pt]{article}
\usepackage{amssymb, amsmath, amsthm, natbib}
\usepackage{setspace}
\usepackage{hyperref}
\usepackage{fullpage}
\usepackage{graphicx}
\graphicspath{ {./images/} }
\usepackage{caption}
\usepackage{subcaption}
\usepackage{float}
\newtheorem{lemma}{Lemma} 
\newtheorem{proposition}{Proposition}

\newtheorem{corollary}{Corollary}

\newtheorem{claim}{Claim}

\newcommand{\commentout}[1]{}

\title{Preference Learning in School Choice Problems\thanks{We thank Yuichi Imai for excellent research assistance and Mariagiovanna Baccara, Aaron Bodoh-Creed, Anqi Li, Leeat Yariv, and seminar participants for valuable comments.}}
\author{SangMok Lee\thanks{Department of Economics, Washington University in St. Louis, Email:
sangmoklee@wustl.edu.}
\date{\today}}

\begin{document}
\maketitle

\abstract{In school choice, students make decisions based on their expectations of particular schools' suitability, and the decision to gather information about schools is influenced by the acceptance odds determined by the mechanism in place. We study a school choice model where students can obtain information about their preferences by incurring a cost. We demonstrate greater homogeneity in rank-order reports and reduced information acquisition under the Deferred-Acceptance (DA) mechanism, resulting in an increased reliance on random tie-breaking and ultimately inefficient outcomes. Thus, it is critical for the DA mechanism to have easy access to school information in order to maintain its efficiency.}

\section{Introduction}

The field of market design in public school choice has grown significantly since the publication of \cite{abdulkadirouglu2003school}. The standard approach in the literature is to adopt \cite{gale1962college}'s two-sided matching model and view students and schools as agents with preferences over potential partners. In practice, school districts assign students to schools based on students' preference reports and schools' priorities, which may consider factors such as test scores and proximity.

The two most prominent mechanisms are the Immediate-Acceptance (also known as the ``Boston'' mechanism) and the Deferred-Acceptance (DA) mechanisms. Both of these mechanisms use student preference reports and school priorities to run a series of algorithm rounds. In each round, students apply for their top-choice schools. If a student is rejected by their first choice, they move on to their next choice and apply again in the next round. The Boston mechanism considers the matches made in each round to be permanent and starts each new round with unmatched students and unoccupied schools. This means that students who first apply to competitive schools take the risk of losing the opportunity to be matched with their next preferred schools if they are rejected by their top choices. On the other hand, the DA mechanism defers the matches until the end of all rounds. This allows students who have been rejected by their most preferred schools to move down their rank-order lists and potentially displace  previously matched students. This deferred acceptance incentivizes students to provide truthful preferences and makes the DA mechanism strategy-proof.

The market design literature widely supports the use of the Deferred-Acceptance (DA) mechanism due to its strategy-proofness and fairness (also known as no justified envy, or stability).\footnote{This means that a school that a student prefers over the assigned match is only assigned to other students who have higher priorities for that school.} However, the DA mechanism may lead to efficiency loss, particularly when students have similar preferences. This is because students who are near indifferent between schools may try a more competitive school first, forcing other students who have a strong desire for that school to be excluded. This issue has been investigated in various studies such as \cite{miralles2009school}, \cite{abdulkadirouglu2011resolving}, \cite{pathak2008leveling}, and \cite{featherstone2011school}.

In our view, what is commonly referred to as preferences in the standard model are better understood as expectations. This is because families do not have firsthand experience with different schools, as they do not regularly consume or purchase them. To form an idea of what their actual preferences might be, families gather information from various sources, such as school district websites, parent forums, and review and rating websites such as GreatSchools.org, niche.com, or SchoolDigger.com. The aggregation of available information -- e.g., reading brochures, organizing internet forums, and visiting open houses -- into application decisions consumes a great deal of energy and time in reality, so students have at best interim expectations of actual preferences that reflect the perceived suitability of different schools.

The strength of the DA mechanism -- strategy-proofness and fairness -- remains intact because students continue to have truth-telling incentives and no justified envy with regard to interim expected preferences. However, it is important to reassess the mechanism's efficiency loss. Students and their families decide how much information to gather and which aspects of the schools to investigate based on the expected quality and the admission odds of the schools, considering the cost of obtaining this information. For instance, they may opt to not seek information about a school or only gather information about a school's exceptional fit if the admission chance is low. Accordingly, the decisions are influenced by the mechanism in use and the learning and reporting strategies of other students. Thus, it is inaccurate to discuss the inefficiency of the DA mechanism on the basis of interim expected preferences, when these expected preferences are influenced by the mechanism in use.

We investigate a centralized school choice model where students incur a cost to gather information about their ex-post preferences. We compare the efficiency loss of the Deferred Acceptance (DA) mechanism with that of the Boston mechanism, taking into account that any fixed mechanism creates a game in which students must make decisions about their learning and reporting strategies. Our analysis highlights the potential efficiency loss of the DA mechanism due to reduced information acquisition.\footnote{At first glance, it might seem trivial that students acquire more information under the Boston mechanism because applying for a competitive school first entails the risk of multiple rejections. However, this same risk can also discourage students from gathering information, since they may choose not to apply to these highly competitive schools.}

We build a school choice model that is simple yet sufficient to demonstrate the potential efficiency loss of the DA mechanism. The model consists of a population of a unit mass of infinitesimal students and three schools: $s$ (superior in expectation), $a$ (average), and $b$ (below average). The match payoffs for students are $u_a = 1$, $u_b = 0$, and $u_s = v + \theta$, respectively. While $v > 1/2$ is fixed across students, $\theta$ is an idiosyncratic preference shock distributed uniformly over $[0,1]$ independently across students. For the purposes of illustration, let us assume that each school has a capacity of $1/3$. A given mechanism, Boston or DA, defines a Bayesian game in which each student first acquires information about her unobservable preference shock and then submits a rank-order list. We adopt the Rational Inattention (RI) information acquisition for information gathering, first introduced by \citet{sims1998} and \citet{sims2003}. The RI model allows for adaptability in information acquisition, enabling students and families to select not only the quantity of information they acquire but also the type of information. The value of different forms of information can vary based on the mechanism, and students and families must prioritize the information that is most critical to them to reduce the cost of information acquisition.

First, we analyze a benchmark model in which students can obtain information about their preference shocks at no cost, i.e., can observe their preference shocks. Since $\min\{u_a, u_s\} > u_b$ with probability 1, we assume that students submit either $sab$ or $asb$. A majority of students will report $sab$ rather than $asb$ in either mechanism, as a majority of them prefer school $s$ over $a$ ($Pr[v+\theta > 1] = v > 1/2$). Since the DA mechanism is strategy-proof, assuming there is no indifference ($v + \theta \neq 1$), a student reports $sab$ if and only if $v + \theta > 1$. On the other hand, reporting $sab$ in the Boston mechanism is risky as a failure to match in the first round is likely to trigger another rejection in the next round, resulting in a match with school $b$. Thus, a student will only report $sab$ if their preference type $\theta$ is significantly greater than $1-v$ (\autoref{lem:boston_screen}). On the other hand, the DA mechanism does not discourage near-indifferent students from reporting $sab$, leading to more homogeneous rank-order submissions.  This causes the mechanism to rely more heavily on random tie-breaking in assigning students to schools, resulting in a larger number of near-indifferent students being matched with the in-demand school $s$. As a result, the allocation produced by the DA mechanism is less efficient than that of the Boston mechanism (\autoref{cor:boston_screen}).

Next, suppose that information acquisition is costly. A student's rank-order submission is only partially influenced by her unobservable preference type. In either mechanism, a student is more likely to receive a signal recommending the reporting of $sab$ when her preference type is higher. The optimal signal structure and cost-justifying accuracy of the signals depend on the mechanism and the strategies of other students (\autoref{lem:eq_condition_da}). We find the equilibrium learning and reporting strategies in both the Boston and DA mechanisms (\autoref{prop:equilibrium}). Our equilibrium analysis indicates that the DA mechanism continues to experience efficiency loss compared to the Boston mechanism (\autoref{prop:boston_efficient_info} and \autoref{cor:da_inefficiency_info}), and this loss can become increasingly significant as the marginal cost of obtaining information increases (\autoref{prop:monotone_da}).

The DA mechanism's efficiency loss is due to the homogeneity of the rank-order reports. The higher number of students reporting $sab$ in the DA mechanism results in an increased need for random tie-breaking to assign students to schools, leading to greater inefficiency. 

The homogeneity of rank-order reports under the DA mechanism is partly driven by the strategy-proofness. A student wants to submit $sab$ when her unobserved preference type $\theta$ is above $1-v$. Hence, when information acquisition is costly, a student's optimal information strategy would focus around $1-v$, learning whether the preference type is above or below that value. If a student's unobservable preference type is close to, but greater than, $1-v$, then they are more likely to report $sab$ than $asb$. On the other hand, in the Boston mechanism, a student with a similar preference type is likely to submit $asb$ because optimal information acquisition in the Boston mechanism focuses on a cutoff point much higher than $1-v$ (\autoref{lem:boston_screen}).

Moreover, the homogeneity of rank-order reports is also intensified by the cost of information acquisition. An increased reliance on random tie-breaking due to the homogeneity of rank-order reports discourages the gathering of costly information. This results in rank-order reports being more influenced by expected match payoffs ($E[u_s]=v+\frac{1}{2} > u_a=1$) and become even more homogeneous (\autoref{prop:transition_by_alpha}). This creates a self-reinforcing cycle between homogeneous rank-order reports and reduced information acquisition, further exacerbating the loss of efficiency in the DA mechanism. A lower marginal cost of information acquisition can mitigate the reinforcing cycle and ultimately the efficiency loss by the DA mechanism.

Finally, we would like to note that the school choice model in this study is stylized and not intended to accurately reflect actual school choice environments. The purpose of using this model is to demonstrate the potential for efficiency loss in the DA mechanism. It is worth mentioning that the DA mechanism is not necessarily less efficient than the Boston mechanism in more complex school choice environments, as indicated by studies such as \citep{troyan2012comparing, calsamiglia2014catchment}.\footnote{If different top schools prioritize students differently, such as based on neighborhood criteria, then the fear of being matched with a subpar school through the Boston mechanism may push these students towards their safer options, such as their neighborhood schools, rather than exploring potential exchanges that would enhance their welfare.} While a comprehensive school choice environment is desirable, our focus is to create a simple one to clearly demonstrate the logic behind the efficiency loss of the DA mechanism. The possibility of efficiency loss is likely to persist and play a role in the design of real-life school choice programs.

\subsection{Related Literature}

\citet{chen2017information} also investigates the acquisition of information in the context of school choice and compares the welfare outcomes under different mechanisms. \citet{chen2017information}'s study considers a general school choice model with preference learning, in which students first choose to learn their ordinal preferences and then their cardinal utilities, each at a cost. In contrast, our approach focuses on a simpler school choice environment to demonstrate the efficiency loss in the DA mechanism. The simplicity of our model enables us to consider a flexible information acquisition model and offers comparative statistics through the use of a single information cost parameter. 

Other studies have also explored the significance of preference learning in matching. \citet{lo2018designing} examines the design of a stable outcome by considering a process in which each student sequentially selects a school to investigate, and defines stability in terms of matching outcomes and students' beliefs about preferences. \citet{bade2015serial} considers the scenario of heterogeneous learning costs among students and finds that prioritizing agents with higher information costs through a serial dictatorship approach can lead to increased information acquisition and enhanced efficiency. \citet{harless2018learning} examines the student's choice of which school to learn about, while \citet{kloosterman2018school} investigates the learning process of each student regarding others' preferences.

We study a mechanism design problem with RI agents, emphasizing how the mechanism being used affects the qualitative and quantitative natures of the information acquired by agents. Other papers sharing this scheme include, but are not limited to: \citet{yang2019} and \citet{liyang} on contract designs with an RI agent; and \citet{bloedelsegal} on information design with an RI agent. We use mutual information to measure the cost of information acquisition. \citet{shannon}, \citet{infortheory} and \citet{caplindean} provide information-theoretic, coding-theoretic, and revealed-preference foundations for this modeling choice, respectively. The growing literature on RI is recently surveyed by \citet{caplinsurvey} and \citet{risurvey}.

\section{Model}

A unit mass of students must be assigned to three schools: $s$, which is expected to be superior, $a$, which is average, and $b$, which is below average. Each school $j$ has a capacity $\lambda_j > 0$ such that $\sum_{j\in\{s, a, b\}} \lambda_j = 1$. The preferences of the students are represented by cardinal utilities, with $u_s = v + \theta$, $u_a = 1$, and $u_b = 0$. The constant $v \in (0,1)$ is common to all students, while $\theta$ is a random variable that is uniformly distributed on $[0,1]$, and is independently determined for each student. The realization of $\theta$ captures each student's individual preference shock between schools $s$ and $a$. A student with $\theta > 1-v$ would have a higher match payoff from attending school $s$ compared to school $a$, and the value of $v$ represents the fraction of students with such ex-post preferences. Regardless of the realization of preference shocks, school $b$ is always the least preferred option.

A matching mechanism requests each student to provide a rank-order list that is either $sab$ or $asb$. School $b$ is always placed at the bottom of the list because it is the least preferred match. The mechanism then assigns students to schools based on the populations of students who report $sab$ and $asb$, while respecting the capacity constraint. We consider the Immediate-Acceptance (also known as the Boston) mechanism and the Deferred-Acceptance (DA) mechanism, which we will explain in more detail later.

Before participating in a given mechanism, students can acquire information about their unobservable preference shocks $\theta$ at a cost. The signal structure $\Pi: [0,1] \rightarrow \Delta \mathcal{Z}$ captures the information acquired by a student. Here, $\Pi\left(\cdot \mid \theta\right)$, $\forall \theta \in \left[0,1\right]$, specifies a probability distribution over a finite set $\mathcal{Z}$ of signal realizations conditional on the true preference shock being $\theta$. For each signal realization $z \in \mathcal{Z}$, the student submits a rank-order list that determines her matching outcome. The cost of information acquisition is $\mu \cdot I\left(\Pi\right)$, where $\mu \geq 0$ is a marginal information cost, and $I\left(\Pi\right)$ is the mutual information between the preference shock and the signals generated by $\Pi$. Specifically, $I\left(\Pi\right) =H\left(\theta\right)-\mathbb{E}_{\Pi}\left[H\left(\theta \mid z\right)\right]$, where $H(\cdot)$ denotes the Shannon entropy of a random variable.

It is without loss of generality to consider signal structures that recommend students to submit either $sab$ or $asb$, and students strictly follow this recommendation \citep{matejka2015}.\footnote{The focus on finite signals is not restrictive. The signal structure can be simplified by merging signal realizations that induce $sab$ into a single recommendation and those that induce $asb$ into another recommendation. This results in the same matching payoffs but with lower acquisition costs since the new signal structure is less Blackwell informative. Students must strictly prefer to follow the recommendations they receive, and if they are indifferent, merging the recommendations still saves on the cost of acquiring information.}
Accordingly, we represent the signal acquired by a student by an integrable function $m: [0,1] \rightarrow [0,1]$, where each $m\left(\theta\right)$ specifies the probability that the student is recommended to submit $sab$ when her unobservable preference shock is $\theta \in [0,1]$. Define $\overline{m} \equiv \int_0^1 m (\theta)d\theta$ as the average probability of the recommendation to submit $sab$. The information cost is then $\mu\cdot I(m)$, where $\mu\geq 0$ and 
\begin{align}
I(m) =& \int_0^1 [m(\theta) \ln m(\theta) + (1-m(\theta))\ln (1-m(\theta)) ]d\theta \notag\\
&-\overline{m} \ln \overline{m} - (1-\overline{m})\ln(1-\overline{m}). \label{eqn:shannon_entropy}
\end{align}

\subsection{Matching mechanisms}
\label{sec_discussion_mechanism}
We examine two mechanisms: Immediate-Acceptance (also known as Boston) and Deferred-Acceptance (DA). We assume a single tie-breaking method where student priorities are randomly assigned from a uniform distribution over $[0,1]$ at the beginning of each mechanism and are used by all schools. A student with a higher priority score is ranked higher, and 
students do not observe their priorities when submitting rank-order lists.

Boston mechanism assigns students to schools in multiple rounds. In each round, unmatched students apply to top-choice schools that have not rejected them, and schools keep accepting students based on their priorities until their capacities are reached. The rounds go as follows in our school choice setup. Let $r \in [0,1]$ denote the proportion of students who submit $sab$, and suppose the fraction is large $r \geq 1-\lambda_a$. In the first round, school $s$ receives more applications than its capacity ($r \geq 1-\lambda_a > \lambda_s$), so it accepts $\lambda_s$ students with the highest priorities out of the $r$ students and rejects the remaining $r-\lambda_s$. Meanwhile, all students who submitted $asb$ match with school $a$. School $s$ has reached its capacity, whereas school $a$ still has openings $\lambda_a-(1-r)$. In the second round, students who submitted $sab$ but were rejected by school $s$ apply to school $a$. Only $\lambda_a-(1-r)$ students with the highest priorities are accepted when school $a$ has reached its capacity. The remaining students match with $b$ in the third round. Table \ref{table:boston_mechanism} reports the final allocation of students to schools, including other cases whose derivations are omitted for brevity.
\begin{table}[H]
\small
\centering
 \begin{subtable}[H]{0.3\textwidth}
	\begin{tabular}{c|ccc}& s & a & b\\ \hline
	sab & $\lambda_s$ & $\lambda_a - (1-r)$ & $\lambda_b$ \\
	asb & 0 & $1-r$ & $0$ 
	\end{tabular}  
	\caption{if $r \geq 1-\lambda_a$}
 \end{subtable} \quad
 \begin{subtable}[H]{0.3\textwidth}
	\begin{tabular}{c|ccc}
	& s & a & b\\ \hline
	sab & $\lambda_s$ & $0$ & $r - \lambda_s$ \\
	asb & 0 & $\lambda_a$ & $(1-r)-\lambda_a$ 
	\end{tabular} 
	\caption{if $\lambda_s \leq r \leq 1-\lambda_a$}
 \end{subtable} \quad
 \begin{subtable}[H]{0.3\textwidth}
	\begin{tabular}{c|ccc}
	& s & a & b\\ \hline
	sab & $r$ & $0$ & $0$ \\
	asb & $\lambda_s -r$ & $\lambda_a$ & $\lambda_b$ 
	\end{tabular} 
	\caption{if $r \leq \lambda_s$}
 \end{subtable}
\caption{The final allocation of students to schools under the Boston mechanism. Each panel shows the number of students who submitted the rank-order lists in the row and were matched with the schools in the column.}
\label{table:boston_mechanism}
\end{table}

The DA mechanism is similar to the Boston mechanism, but it defers the assignments in each round until the end of all rounds. The mechanism is also characterized by market-clearing priority level cutoffs \citep{azevedo2016supply}. Market-clearing priority cutoffs $p_s, p_a, p_b \in [0,1]$ are such that assigning each student to her top choice among the schools that have cutoffs lower than the student's priority ranking exactly meets the schools' capacities. It is known that market-clearing cutoffs uniquely exist. In our setup, every student is guaranteed to match at least with school $b$, so $p_b=0$. If a large fraction $r \geq \frac{\lambda_s}{\lambda_s + \lambda_a}$ of students report $sab$, that is, school $s$ receives more applications than school $a$ relative to their capacities, then the market-clearing cutoffs must satisfy $p_s \geq p_a > p_b = 0$.

The assignment of students through the DA mechanism is determined as follows. For students who report $sab$, those whose priority rankings are higher than $p_s$ are assigned to school $s$, while those with priority rankings between $p_a$ and $p_s$ or below $p_a$ are assigned to school $a$ or $b$, respectively. For students who report $asb$, their assignments to either school $a$ or $b$ depend on whether their priority rankings are above or below the cutoff $p_a$. This assignment by priority cutoffs is:
\begin{center}
\begin{tabular}{c|ccc}
 & s & a & b \\ \hline
sab & $(1-p_s)r$ & $(p_s-p_a)r$ & $p_a r$\\
asb & 0 & $(1-p_a)(1-r)$ & $p_a (1-r)$
\end{tabular}.
\end{center}
The cutoffs are market-clearing if $(1-p_s)r = \lambda_s$ and $(p_s - p_a)r + (1-p_a)(1-r) = \lambda_a$, subject to the constraint $p_s \geq p_a > p_b=0$. By replacing the priority cutoffs with market-clearing values, we obtain the student assignment, which is shown in Panel (a) of Table \ref{table:DA_mechanism}. The other case is not shown due to brevity.\footnote{\label{rmk:ttc}The Top-Trading-Cycles (TTC) mechanism with a random endowment is equivalent to the DA in our setting. For any proportion $r$ of students who report $sab$, the initial assignment of students is
\begin{center}
\begin{tabular}{c|ccc}
& s & a & b \\ \hline
sab & $r\lambda_s$ & $r\lambda_a$ & $r \lambda_b$\\
asb & $(1-r)\lambda_s$ & $(1-r)\lambda_a$ & $(1-r)\lambda_b$
\end{tabular}.
\end{center}
Then, students who reported $sab$ but are assigned to school $a$ trade with those who reported $asb$ but are assigned to school $s$. The resulting assignment is the same as shown in Panel (a) of Table \ref{table:DA_mechanism}. For a general environment with a continuum of students, see \citet{leshno2017simple} for a characterization of the Top-Trading-Cycles (TTC) mechanism.}
\begin{table}[H]
\centering \small
 \begin{subtable}[H]{0.45\textwidth}
 \centering
 	 \begin{tabular}{c|ccc}
	 & s & a & b \\ \hline
	 sab & $\lambda_s$ & $r\lambda_a - (1-r)\lambda_s$ & $\lambda_b r$\\
	 asb & 0 & $(1-r)(\lambda_s + \lambda_a)$ & $\lambda_b (1-r)$
	 \end{tabular}
	\caption{if $r \geq \frac{\lambda_s}{\lambda_s + \lambda_a}$}
 \end{subtable}\quad 
 \begin{subtable}[H]{0.45\textwidth}
 \centering
	\begin{tabular}{c|ccc}
	& s & a & b \\ \hline
	sab & $r(\lambda_s + \lambda_a)$ & $0$ & $r \lambda_b$\\
	asb & $(1-r)\lambda_s - r \lambda_a$ & $\lambda_a$ & $(1-r)\lambda_b$
	\end{tabular}
	\caption{if $r < \frac{\lambda_s}{\lambda_s + \lambda_a}$}
 \end{subtable}
\caption{The final allocation of students to schools under the DA mechanism. Each panel shows the number of students who submitted the rank-order list in the row and were matched with the schools in the column.}
\label{table:DA_mechanism}
\end{table}

\subsection{Information acquisition}
Each mechanism $\Gamma \in \{\text{Boston}, \text{DA}\}$ defines a game in which a student $i$ must incur a cost to acquire information about her unobservable preference shock $\theta$. After acquiring information, the student submits a rank-order list $sab$ or $asb$ to the mechanism.

A student's problem is defined as follows. For any given mechanism $\Gamma$ and a fraction $r$ of students that report $sab$, $U^{\Gamma}_{sab}(\theta; r)$ and $U^{\Gamma}_{asb}(\theta;r)$ denote the expected match payoffs that a student with a preference type $\theta$ obtains by choosing a rank-order list $sab$ and $asb$, respectively. Let $\Delta^{\Gamma}(\theta;r) \equiv U^{\Gamma}_{sab}(\theta; r)-U^{\Gamma}_{asb}(\theta;r)$. Then, a student solves
\begin{equation}\label{eqn_problem}
\max_{m: [0,1] \rightarrow [0,1]} \int_0^1 m(\theta) \Delta^{\Gamma}(\theta ; r) d\theta - \mu I(m).
\end{equation}

\begin{figure}[h!]
\centering
 \begin{subfigure}{.48\textwidth}
 \centering
 \includegraphics[scale=0.25]{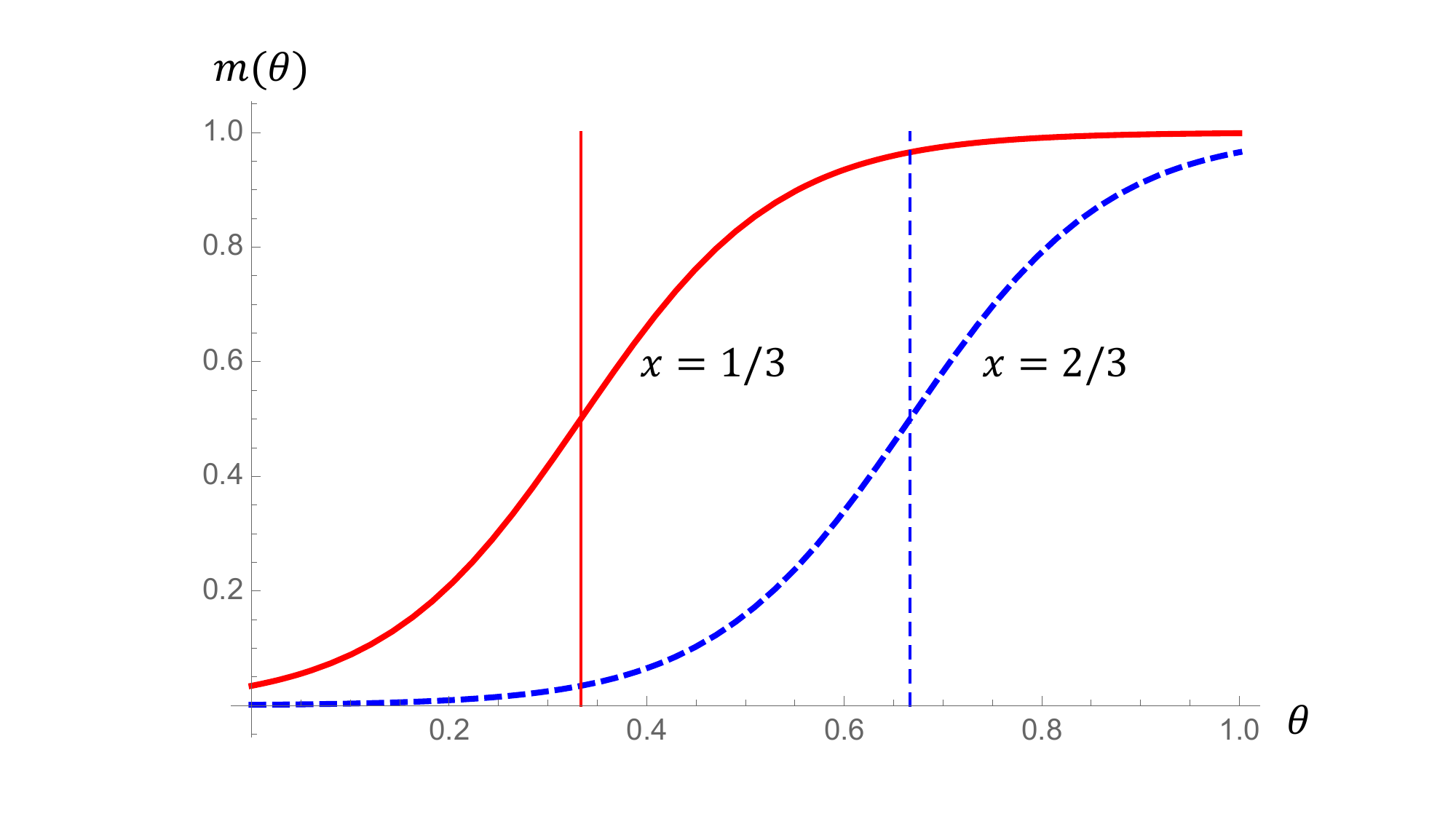}
 \caption{A decision of what to learn ($\mu = 0.1$).}\label{fig:info_ac_a}
 \end{subfigure} 
 \begin{subfigure}{.48\textwidth}
 \centering
 \includegraphics[scale=0.25]{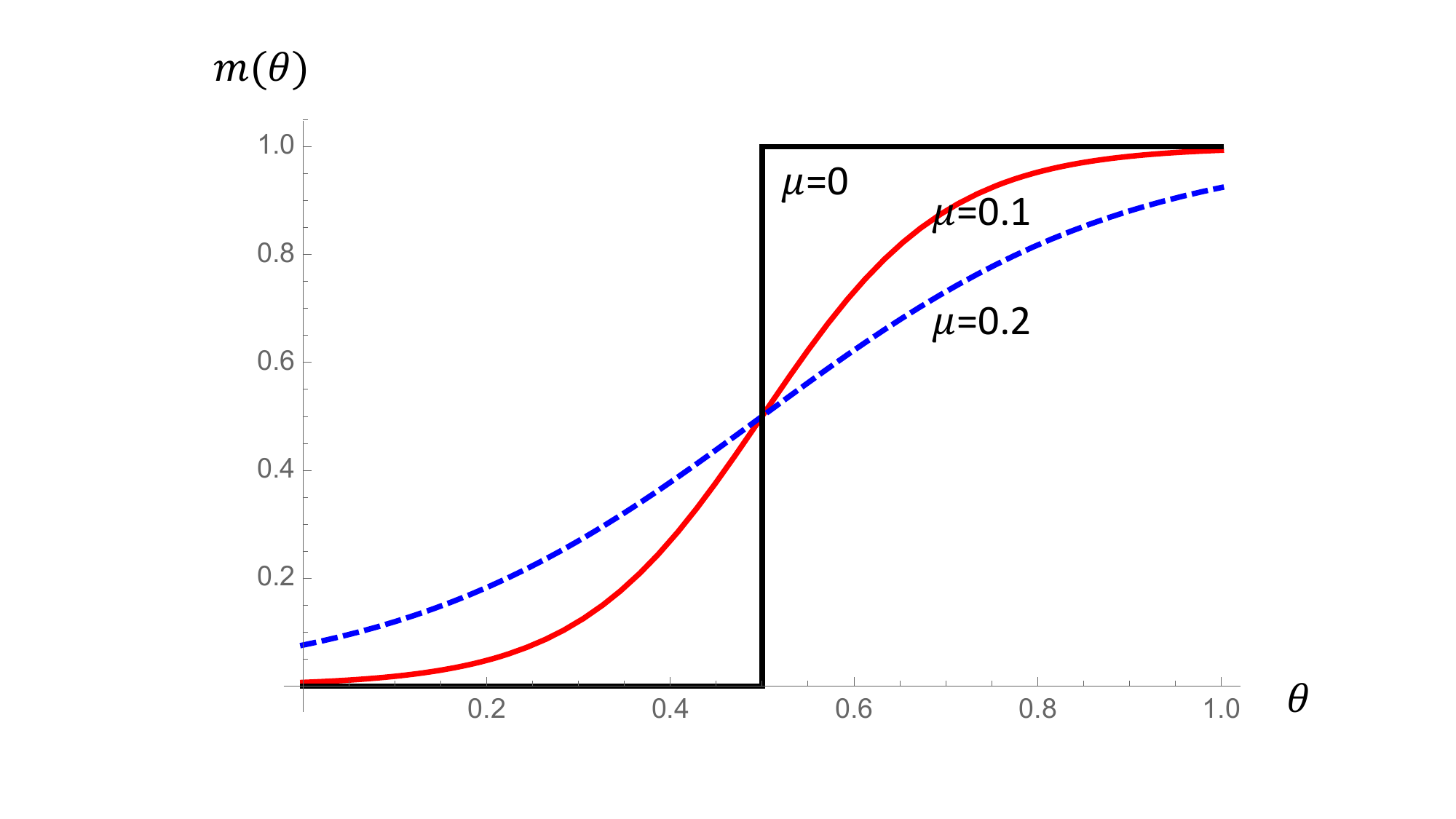}
 \caption{A decision of how much to learn ($x=1/2$).}\label{fig:info_ac_b}
 \end{subfigure}
\caption{An optimal information acquisition strategy when $\Delta(\theta) = \theta - x$.}
\label{fig:info_acquisition}
\end{figure}

An optimal signal structure represents the optimal decision of what and how much a student should learn about her preference type. In \autoref{fig:info_acquisition}, we illustrate an optimal signal structure when $\Delta(\theta) = \theta - x$ for $x \in (0,1)$.
The optimal signal structure $m(\theta)$ increases with $\theta$ because the payoff gain $\Delta(\theta)$ by switching from $asb$ to $sab$ increases with $\theta$. The agent aims to choose $sab$ if and only if her preference type $\theta$ is above $x$. Therefore, the agent chooses to learn whether her preference shock $\theta$ is above or below the option value $x$, as shown in Panel (a). If the marginal cost of information $\mu$ decreases, the agent chooses to acquire more information, resulting in the recommendation by $m(\theta)$ responding more precisely to $\theta$, as shown in Panel (b).

\section{Equilibrium Analysis}

A game is defined by any fixed mechanism $\Gamma \in \{\text{Boston}, \text{DA}\}$. We find a symmetric Nash equilibrium $m^{\Gamma}$ in this game. The proportion of students reporting $sab$ in $m^{\Gamma}$ is $r^{\Gamma} \equiv \int m^{\Gamma}(\theta)d\theta$. Conversely, $m^{\Gamma}$ is the unique solution to the information acquisition problem (\autoref{eqn_problem}) for each student given the fraction $r^{\Gamma}$ of other students who report $sab$. Therefore, it is sufficient to find an equilibrium proportion $r^{\Gamma}$ of students reporting $sab$.

Our focus is on parameter values that lead to a higher proportion of students submitting $sab$ than $asb$ in equilibrium, relative to the schools' capacities. Specifically, given a set of capacity profiles $(\lambda_s, \lambda_a, \lambda_b)$ and marginal information cost $\mu$, we assume a sufficiently high constant $v$ to ensure that the equilibrium fraction of students submitting $sab$ is greater than $\hat{r} \equiv \frac{\lambda_s}{\lambda_s+\lambda_a}$ in both mechanisms.\footnote{When the marginal cost of information $\mu$ is very high, assuming $v > \frac{1}{2}$ is sufficient because most students will choose $sab$ based on the ex-ante utilities, $E[u_s] = v + E[\theta] > u_a$. However, if the marginal cost $\mu$ is low, the condition for $v$ depends on the schools' capacities. A school with higher quality for the majority of students, $Pr[v + \theta > u_a] = v > \frac{1}{2}$, can be less selective if its capacity is very large.} Consequently, in equilibrium, school $s$ is more competitive than school $a$: the Boston mechanism fills up school $s$ no later than school $a$, and under the DA mechanism, the priority cutoff for school $s$ is weakly higher than for school $a$.

For the Boston mechanism, we consider two cases: $r \geq 1-\lambda_a$ (Panel (a) of \autoref{table:boston_mechanism}) and $\hat{r} < r < 1-\lambda_a$ (Panel (b) of \autoref{table:boston_mechanism}). When $r \geq 1-\lambda_a$, if a student submits $sab$, the mechanism assigns her to schools $s$, $a$, and $b$, with probabilities $\frac{\lambda_s}{r}$, $\frac{\lambda_a - (1-r)}{r}$, and $\frac{\lambda_b}{r}$, respectively. If a student submits $asb$, the mechanism assigns her to school $a$. We conduct a similar analysis for the case of $\hat{r} < r < 1-\lambda_a$. The expected gain in match payoffs $\Delta^{B}(\theta;r)$ for any fraction $r > \hat{r}$ of students who submit $sab$, given an unobservable preference type $\theta$, is
\begin{equation}
\Delta^{B}(\theta; r) = \frac{\lambda_s (v+ \theta)}{r} - \min\left\{\frac{\lambda_a}{1-r}, \frac{1-\lambda_a}{r}\right\}, \quad \forall r > \hat{r}.
\label{eqn:delta_boston}
\end{equation}

For the DA mechanism (Panel (a) of \autoref{table:DA_mechanism}), if a student reports $sab$, the mechanism assigns her to schools $s$, $a$, and $b$, with probabilities $\frac{\lambda_s}{r}$, $\lambda_a - \frac{(1-r)\lambda_s}{r}$, and $\lambda_b$, respectively. If the student reports $asb$ instead, the probabilities become $0$, $\lambda_s + \lambda_a$, and $\lambda_b$. This implies that
\begin{align} 
\Delta^{D}(\theta; r) & = \left(
\frac{\lambda_s}{r} (v + \theta) + \lambda_a - \frac{1-r}{r}\lambda_s\right) - \left(\lambda_s + \lambda_a\right)& \notag\\
& = \frac{\lambda_s(v + \theta)}{r} - \frac{\lambda_s}{r},& \forall r > \hat{r}.
\label{eqn:delta_da}
\end{align}

\begin{lemma}
\label{lem:boston_avoid_competition}
For any $r > \hat{r}$ and $\theta \in [0,1]$, 
$\Delta^{D}(\theta; r) > \Delta^{B}(\theta; r).$
\end{lemma}
\begin{proof}
$\Delta^{D} - \Delta^{B}=\min\left\{\frac{\lambda_a}{1-r}, \frac{1-\lambda_a}{r} \right\}- \frac{\lambda_s}{r}$. Clearly, $\frac{1-\lambda_a}{r} = \frac{\lambda_b + \lambda_s}{r} > \frac{\lambda_s}{r}$. Moreover, $r > \hat{r}$ implies $\frac{\lambda_a}{1-r} > \frac{\lambda_s}{r}$.
\end{proof}

The intuition behind \autoref{lem:boston_avoid_competition} is that when a large proportion of students apply to school $s$ first ($r > \hat{r}$) in both mechanisms, submitting $sab$ to the Boston mechanism involves a higher risk for a student. This is because failing to match with the top-choice school $s$ in the first round is likely to lead to another rejection in the second round, potentially resulting in a final match with the worst school $b$. In contrast, under the DA mechanism, the risk of matching with school $b$ does not increase from choosing a rank-order list $sab$.

\subsection{A benchmark case of free information acquisition.}
We study the case of free information acquisition, where $\mu=0$. This corresponds to a standard model where a student knows her own preferences. For any mechanism $\Gamma \in \{\text{Boston}, \text{DA}\}$, an equilibrium strategy must be
$$
m^{\Gamma}(\theta)= \left\{\begin{array}{ll}
1 & \text{if $\theta > \theta^{\Gamma}$} \\
0 & \text{if $\theta < \theta^{\Gamma}$}.
\end{array}\right.
$$
with a mechanism-dependent threshold $\theta^{\Gamma}$. We omit the value at threshold $m^{\Gamma}(\theta^{\Gamma})$ since any value in $[0,1]$ is consistent with our equilibrium analysis. For the DA mechanism with single tie-breaking, a (strict) dominant strategy for a student is to use the threshold $\theta^D = 1-v$. The proportion of students who report $sab$ is $r^D = 1-\theta^D = v$, which is assumed to be larger than $\hat{r} = \frac{\lambda_s}{\lambda_s + \lambda_a}$. For the Boston mechanism, the equilibrium threshold $\theta^B$ is the unique solution of $\Delta^{B}(\theta ; r^B) = 0$, where $r^B = 1-\theta > \hat{r}$.\footnote{A unique solution $\theta^B$ satisfies either (i) $\theta^{B} \leq \lambda_a$ and $\lambda_s (v+\theta^{B}) = 1-\lambda_a$, or (ii) $\lambda_a \leq \theta^{B} \leq 1-\lambda_s$ and $\frac{\lambda_s (v+\theta^{B})}{1-\theta^{B}} = \frac{\lambda_a}{\theta^{B}}$.} According to \autoref{lem:boston_avoid_competition}, a student will submit $sab$ only if her preference shock $\theta$ is significantly greater than $\theta^D = 1-v$.
\begin{lemma} If $v > \hat{r}$, then $\theta^{B} > \theta^{D}$. 
\label{lem:boston_screen}
\end{lemma}

\begin{proof}
By \autoref{lem:boston_avoid_competition}, if $r > \hat{r}$, then $\Delta^{B}(\theta^D; r) < \Delta^{D}(\theta^D; r) = 0$, where $\theta^D = 1-v$. Since $\Delta^{B}(\theta; r)$ is strictly increasing in $\theta$, the unique solution $\theta^B$ of $\Delta^B(\theta; r^B)=0$ where $r^B = 1-\theta > \hat{r}$ must satisfy $\theta^B > \theta^D$.
\end{proof}

Under the Boston mechanism, students who are near-indifferent between schools $s$ and $a$ due to preference shocks $\theta \in (\theta^{D}, \theta^{B})$ are deterred from submitting a rank-order list $sab$ due to the higher risk of receiving multiple rejections. However, such students submit $sab$ since the DA mechanism is strategyproof. As a result, in the equilibrium of the DA mechanism, a larger proportion of students choose a more competitive school $s$ as their top choice, resulting in more homogeneous rank order reports. These homogeneous reports cause the DA mechanism to rely on random tie-breaking to a greater extent, which can lead to an efficiency loss.

We define the efficiency of an allocation as follows. Every allocation assigns students to schools $a$ and $b$ up to their capacity, and all students assigned to either school receive the same match payoffs, $u_a = 1$ or $u_b = 0$. Therefore, the efficiency of an allocation can be evaluated based on the students assigned to school $s$. For any mechanism $\Gamma \in {\text{Boston}, \text{DA}}$, a student with a preference shock $\theta$ is assigned to school $s$ with probability $g^{\Gamma}(\theta) \equiv m^\Gamma(\theta)\frac{\lambda_s}{r^\Gamma}$, where $m^\Gamma(\theta)$ represents the probability that a student submits a rank-order list $sab$ given the preference shock $\theta$, and $\frac{\lambda_s}{r^\Gamma}$ represents the probability of matching with school $s$ given that the student submits a rank-order list. In equilibrium, since school $s$ is more competitive than school $a$ ($r^{\Gamma} > \hat{r}$), a student can only match with school $s$ if she submits a rank-order list $sab$. The function $g^{\Gamma}(\theta)$ represents the density of preference shock $\theta$ among students assigned to school $s$ in equilibrium, and it satisfies the condition that $\int g^{\Gamma}(\theta) d\theta = \lambda_s$, where $\lambda_s$ is the number of seats available at school $s$. For each mechanism $\Gamma \in \{\text{Boston}, \text{DA}\}$, the equilibrium allocation is 
$$
g^{\Gamma}(\theta) \equiv \left\{\begin{array}{ll}
\frac{\lambda_s}{1-\theta^{\Gamma}} & \text{if $\theta > \theta^{\Gamma}$,}\\
0 & \text{if $\theta < \theta^{\Gamma}$}.
\end{array}\right.
$$

We say that an allocation $g$ is \emph{more efficient} than allocation $g'$ if $g$ first-order dominates $g'$, that is, $\int_0^{\overline{\theta}} g(\theta) d\theta \leq \int_0^{\overline{\theta}} g'(\theta) d\theta$ for every $\overline{\theta} \in (0,1)$, with strict inequality for at least one $\overline{\theta} \in (0,1)$.

\begin{corollary}
\label{cor:boston_screen}
In the case of free information acquisition, the DA mechanism is less efficient than the Boston mechanism.
\end{corollary}
The proof is straightforward from \autoref{lem:boston_screen}, which establishes that $\theta^B > \theta^D$. Under the Boston mechanism, students with higher preference shocks apply for school $s$, and each such student has a higher chance of matching with school $s$.

\subsection{Equilibrium under costly information acquisition}

We continue to consider parameter values such that, for each mechanism $\Gamma \in \{\text{Boston}, \text{DA}\}$, a symmetric Nash equilibrium $m^{\Gamma}(\theta)$ satisfies $r^{\Gamma} \equiv \int_0^1 m^{\Gamma}(\theta) d\theta \in (\hat{r},1)$. In other words, the value of $v$ is sufficiently high to make school $s$ more competitive than school $a$, but not so high that an interior equilibrium arises.\footnote{We exclude the case of a non-interior equilibrium where $v$ is very close to 1, as in that case, under the DA mechanism, students do not acquire any information, and the equilibrium allocation becomes independent of their preference types.}

To find an interior equilibrium of any fixed mechanism $\Gamma \in \{\text{Boston}, \text{DA}\}$, we use the fact that a student's optimal strategy when a proportion $r^{\Gamma}$ of other students report $sab$ is a solution to the information acquisition problem (\autoref{eqn_problem}). An interior solution satisfies the first-order condition, as shown in \citet{yang2015coordination}:
\begin{equation}
\Delta^{\Gamma} (\theta; r^{\Gamma}) = \mu \cdot \left[\ln \left(\frac{m(\theta)}{1-m(\theta)}\right) - \ln \left(\frac{\overline{m}}{1-\overline{m}}\right)\right], \quad \forall \theta \in [0,1],
\label{eqn:eq_foc}
\end{equation}
where $\overline{m} \equiv \int m(\theta) d\theta \in (0, 1)$. Accordingly, if $m(\cdot)$ is an interior symmetric Nash equilibrium, then by substituting $\overline{m}$ with $r^{\Gamma}$ in \eqref{eqn:eq_foc},
\begin{align}
m(\theta) = \left(1+\frac{1-r^{\Gamma}}{r^{\Gamma}} \exp\left(-\frac{\Delta^{\Gamma}(\theta; r^{\Gamma})}{\mu}\right)\right)^{-1}, \quad \forall \theta \in [0,1].
\label{eqn:m_uniquely_by_r}
\end{align}
Finally, the consistency $r^{\Gamma} = \int m(\theta) d\theta$ implies the following equilibrium condition:
\begin{lemma}\label{lem:eq_condition_da}
An information acquisition strategy $m(\cdot)$ with $\overline{m} \in (\hat{r}, 1)$ is a symmetric Nash equilibrium under a mechanism $\Gamma \in \{\text{Boston}, \text{DA}\}$ if and only if it satisfies \eqref{eqn:m_uniquely_by_r}, where $r^\Gamma \in (\hat{r}, 1)$ is a solution to
\begin{equation}
\exp\left(\frac{\lambda_s}{\mu}\right) = 1 + \frac{\exp\left(\frac{\lambda_s}{r \mu}\right) -1}{\frac{1-r}{r} \exp\left(- \frac{\Delta^{\Gamma}(0 ; r)}{\mu}\right) +1}.
%\exp\left(\frac{\lambda_s}{\mu}\right) = 1 + \frac{\exp\left(\frac{\lambda_s}{r \mu}\right) -1}{\frac{1-r}{r} \exp\left(\frac{\lambda_s}{r\mu}\right)\exp\left(-\frac{\lambda_s v}{r \mu}\right) +1}.
\label{eqn:eq_restriction_da}
\end{equation}
%A solution $r^{\Gamma}$ of \eqref{eqn:eq_restriction_da} uniquely identifies an equilibrium strategy by \eqref{eqn:m_uniquely_by_r}.
\end{lemma}
Therefore, it is sufficient to find a proportion $r^\Gamma \in (\hat{r}, 1)$ of students who submit the rank-order list $sab$ in equilibrium. A careful inspection of \eqref{eqn:eq_restriction_da} yields:
\begin{proposition}
\label{prop:equilibrium}
For any $\mu > 0$,
\begin{enumerate}
\item there exist $\underline{v},\overline{v} \in (0,1)$ with $\underline{v} < \overline{v}$ such that $v \in (\underline{v}, \overline{v})$ if and only if an interior equilibrium of the DA mechanism (uniquely) exists with $r^{D} \in (\hat{r},1)$, and
\item if the DA mechanism has an interior equilibrium with $r^{D} \in (\hat{r}, 1)$, then the Boston mechanism also has a (unique) interior equilibrium with $r^{B} \in (\hat{r}, r^{D})$.
\end{enumerate}
\end{proposition}

The proof of \autoref{prop:equilibrium} is tedious but straightforward. The left-hand side (LHS) of \eqref{eqn:eq_restriction_da} is constant at $\exp(\frac{\lambda_s}{\mu})$. If the DA mechanism is given, the right-hand side (RHS) equals the LHS at $r=1$ and increases indefinitely as $r$ approaches $0$. The RHS is shown to be a strictly single-dipped function.\footnote{A single-dipped function is the $-1$ multiple of a single-peaked function.} The intermediate value theorem implies that a unique solution of \eqref{eqn:eq_restriction_da} exists in $(0,1]$. Then, since the RHS is strictly increasing in $v$ for each fixed $r < 1$, we can find the bounds of $v$ that ensure the solution is in $(\hat{r}, 1)$. For the Boston mechanism, note that $\Delta^{D}(0, r) > \Delta^{B}(0, r)$ for every $r > \hat{r}$ (\autoref{lem:boston_avoid_competition}). So, the RHS of \eqref{eqn:eq_restriction_da} under the Boston mechanism is smaller than under the DA for every $r \in (\hat{r}, 1)$, and the two coincide at $r=\hat{r}$ and $r =1$. It follows that a solution $r^B >\hat{r}$ under the Boston mechanism is smaller than $r^D$.

\subsection{The DA mechanism's efficiency loss}

As mentioned earlier, we measure allocation efficiency by examining the distribution of preference shocks among students assigned to school $s$, denoted by $g^\Gamma(\theta)$ for each mechanism $\Gamma \in \{\text{Boston}, \text{DA}\}$.

\begin{proposition}
\label{prop:boston_efficient_info}
If an equilibrium with $r^{D} \in (\hat{r}, 1)$ exists for the DA mechanism (which implies the existence of equilibrium with $r^B \in (\hat{r}, r^D)$ for the Boston mechanism), then $g^B$ is single-crossing $g^D$ from below.\footnote{This means that for all $\theta' < \theta''$, if $g^B(\theta') \geq g^D(\theta')$, then $g^B(\theta'') > g^B(\theta'')$.} 
\end{proposition}

\begin{corollary}
\label{cor:da_inefficiency_info}
The DA mechanism has a lower allocation efficiency than the Boston mechanism:
$$\int_0^{\overline{\theta}} g^D(\theta) d\theta > \int_0^{\overline{\theta}} g^{B}(\theta) d\theta, \quad \forall \overline{\theta} \in (0,1).$$
\end{corollary}

\begin{figure}[h]
\centering
\includegraphics[scale=0.3]{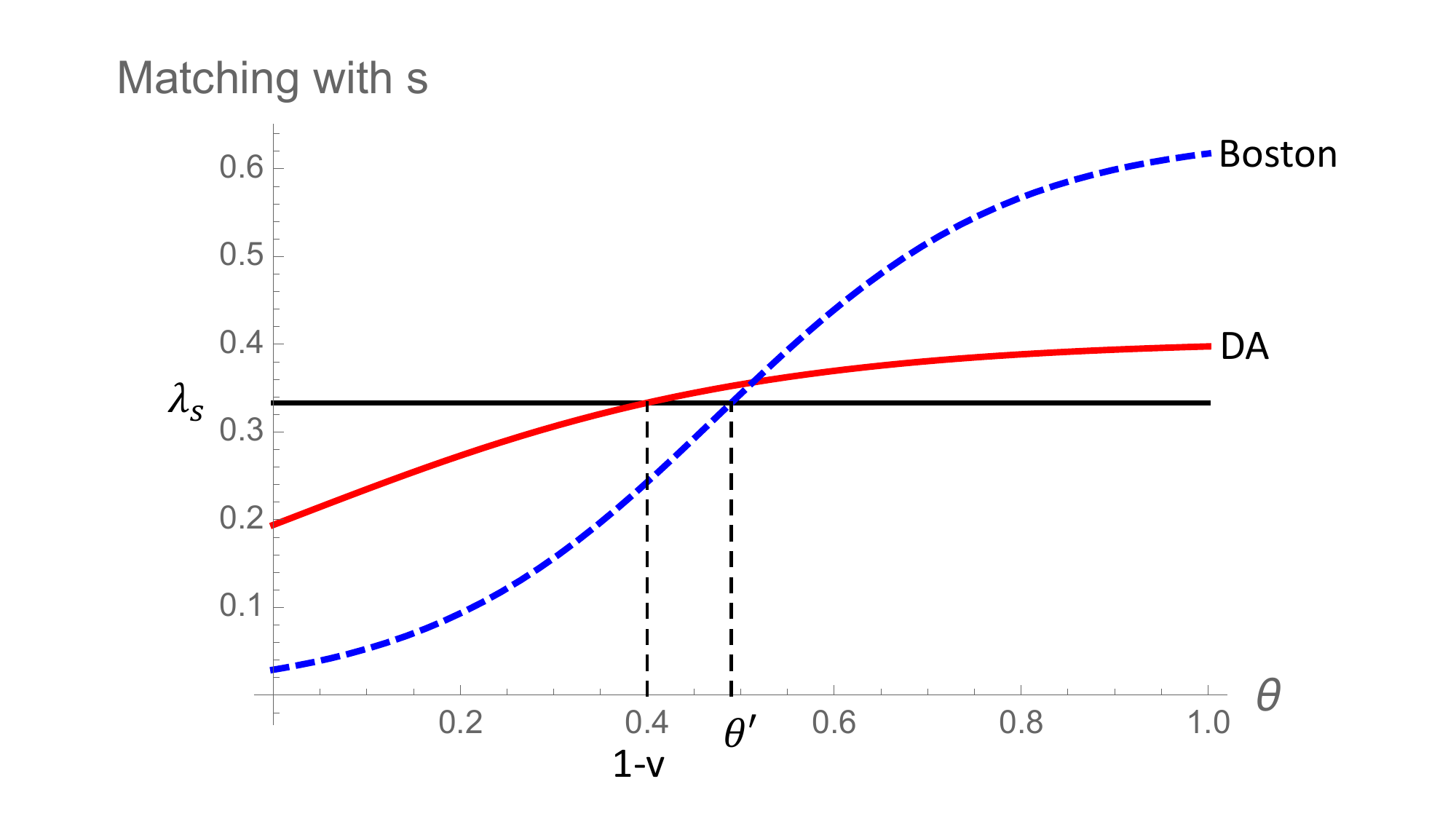}
\caption{The allocation probabilities of matching with school $s$ in the equilibrium of the DA and Boston mechanisms, with parameter values $v = 0.6$, $\mu = 0.1$, and $\lambda_j = 1/3$ for $j \in \{s, a, b\}$.}
\label{fig:eff_comparison}
\end{figure}

\autoref{fig:eff_comparison} illustrates \autoref{prop:boston_efficient_info} and \autoref{cor:da_inefficiency_info} for the parameter values $v=0.6$, $\mu = 0.1$, and $\lambda_j = 1/3$ for $j \in \{s, a, b\}$. The black horizontal line at $\lambda_s$ represents a purely random assignment of students to school $s$. The red solid line represents the allocation of school $s$ in the equilibrium of the DA mechanism ($g^{D}$). The graph intersects the line at $\lambda_s$ when a preference shock causes the match payoff from schools $s$ and $a$ to become identical ($\theta = 1-v$), such that a student is indifferent between reporting $sab$ and $asb$. The blue dashed line represents the allocation of school $s$ in the equilibrium of the Boston mechanism ($g^{B}$). A student who submits $sab$ to the Boston mechanism bears a higher risk of matching badly with school $b$. Therefore, a higher preference shock $\theta' > 1-v$ is required for a student to be indifferent between rank order lists $sab$ and $asb$. The graph of $g^{B}$ is single-crossing $g^D$ from below, which implies the inefficiency of the DA mechanism.

We examine the reasons for the DA mechanism's efficiency loss compared to the Boston mechanism, focusing on the role of homogeneous rank-order reports. The benchmark model, which assumes free information acquisition, showed that students are inclined to submit more uniform rank-order lists to the DA mechanism. This tendency increases the mechanism's dependence on tie-breaking, resulting in reduced disparities in the expected match payoffs across different rank-order choices. As a consequence, students are discouraged from investing in information acquisition when the costs associated with such efforts increase. As students acquire less information, their rank-order reports become less responsive to preference shocks, resulting in greater uniformity. Consequently, the incentives for information acquisition diminish further, exacerbating the cycle.

%Formally, we use the derivative of an information acquisition strategy at $\theta$, i.e., $\frac{dm(\theta)}{d\theta}$, to measure how much information is acquired to distinguish the preference types locally around $\theta$. \begin{proposition} \label{prop:more_info_boston_new} Suppose that the DA and the Boston mechanisms have equilibrium with fractions $r^{D}$ and $r^{B}$ in $(\hat{r}, 1)$. Then,$\frac{dm^B(\theta)}{d\theta}$ is single crossing $\frac{dm^D(\theta)}{d\theta}$ from below.\end{proposition}

To demonstrate the feedback loop between homogeneous rank-order reports and reduced information acquisition, we investigate how students' behavior evolves gradually in response to a switch from the Boston to the DA mechanism. We assume appropriate parameter values to ensure that the equilibrium proportions of students who report the $sab$ list to the DA and Boston mechanisms ($r^D$ and $r^B$) fall within the interval $(\hat{r}, 1)$ (as in  \autoref{prop:equilibrium}). Starting from the Boston mechanism equilibrium, we examine the efficiency of the allocation resulting from a gradual transition to the DA mechanism, with students' beliefs about others' strategies held fixed at $r^B \in (\hat{r}, 1)$. Specifically, for any $r \in [r^B, r^D]$, we define $m^H(\cdot; r, \mu)$ as the students' interior optimal strategy when they believe that an $r$ proportion of their peers will report the $sab$ list to the DA mechanism. We also define $\overline{m}^H(r, \mu) \equiv \int m^H(\theta ; r, \mu) d\theta$. Finally, we measure the efficiency of the resulting allocation using $g^H(\theta ; r, \mu) \equiv m^H(\theta; r, \mu) \frac{\lambda_s}{\overline{m}^H(r, \mu)}$.

To begin, we analyze how students respond to the transition from the Boston to DA mechanism, while holding their belief about others' strategies fixed at $r^B$. 
\begin{proposition}
\label{prop:transition_by_beta}
Take $v \in (\underline{v}, \overline{v})$ such that the DA and the Boston mechanisms have interior equilibria with $\hat{r} < r^B < r^D < 1$. Let $m^H(\cdot ; r^B, \mu)$ be the optimal strategy when the mechanism changes from Boston to DA while the students' belief on the proportion of others reporting $sab$ is fixed $r^B$. Then, (i) $\overline{m}^H(r^{B}, \mu) > r^B$, and (ii) $g^{B}$ is single crossing $g^H(\cdot; r^{B}, \mu)$ from below.
\end{proposition}

To understand the rationale behind \autoref{prop:transition_by_beta}, consider the mechanism transitions from Boston to DA, while the students' belief about others' strategies is fixed at $r^B$. This fixed belief means that the chance of matching with school $s$ given the choice of rank-order report $sab$ is believed to remain unchanged. However, the switch to the DA mechanism eliminates the risk of a bad match resulting from submitting $sab$.

Next, we study the subsequent adjustments in the students' strategy.
\begin{proposition} \label{prop:transition_by_alpha}
Suppose $\mu < \lambda_s (1-v)$. If $r_1, r_2 \in (r^B, r^D)$ with $r_1 < r_2$, then $r^B < \overline{m}^H(r_1, \mu) < \overline{m}^H(r_2, \mu) < r^D$, and the allocation $g^H(\cdot ; r_1, \mu)$ is single-crossing $g^H(\cdot; r_2, \mu)$ from below.
\end{proposition}
The assumption $\mu < \lambda_s (1-v)$ is to ensure that an interior optimal strategy under the DA mechanism $m^H(r, \mu)$ exists for any $r \in (\hat{r}, 1)$.

For the first part of \autoref{prop:transition_by_alpha}, note that the given mechanism is DA, but the belief on others' strategies $r$ is greater than $r^D$. Since the rank-order choices are believed to be more homogeneous ($r > r^D > \hat{r}$), the mechanism is believed to rely on random tie-breaking to a greater degree. In response, students acquire less information, and their rank-order reports become less sensitive to preference shocks and more homogeneous, leading to a reduction in the incentives for information acquisition. A similar intuition applies to the second part.

\begin{figure}[h]
\centering
\includegraphics[scale=0.4]{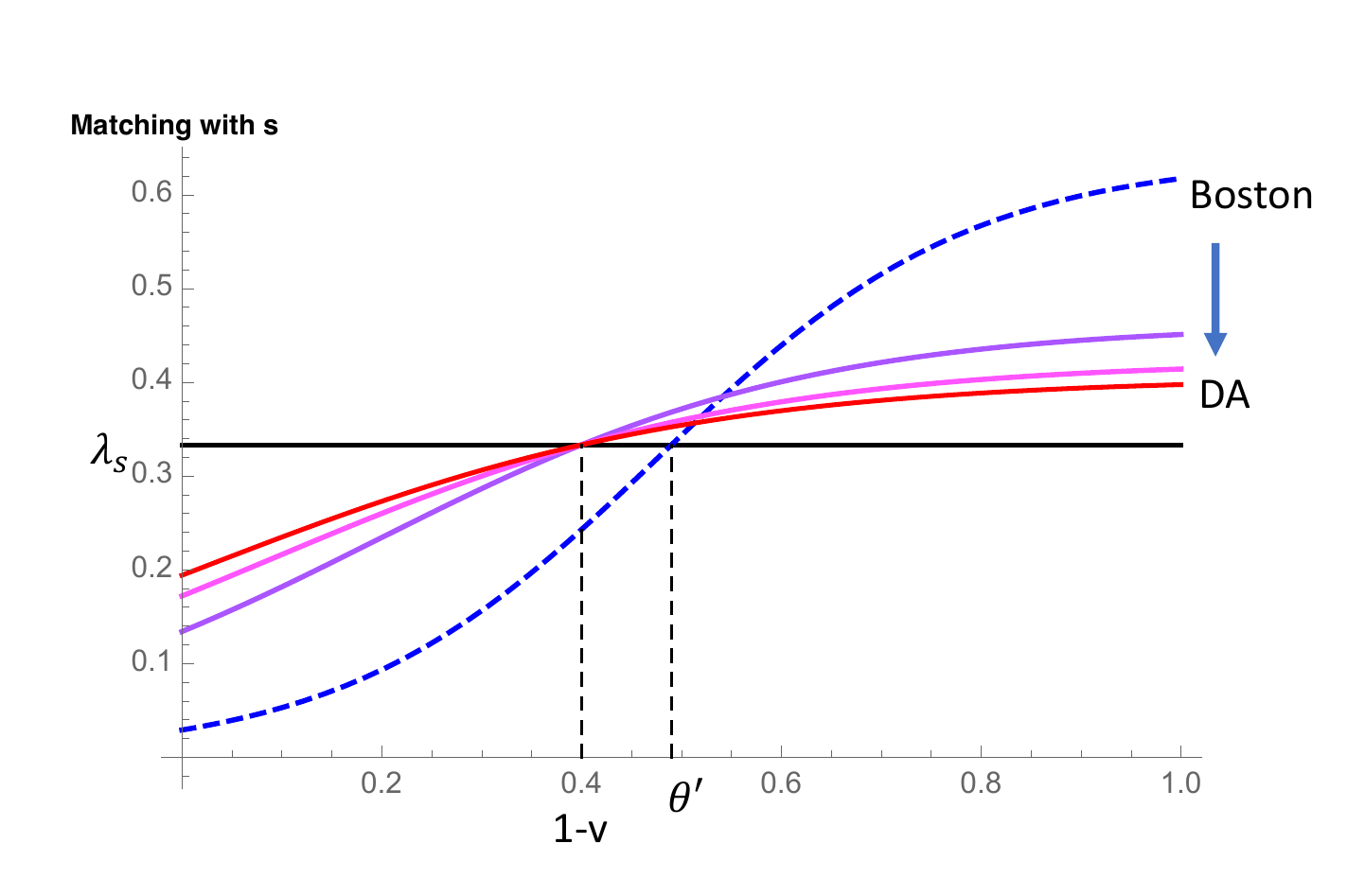}
\caption{Allocation efficiency declines after a switch from the Boston mechanism to the DA followed by best-response adjustments, with parameter values of $v=0.6$, $\mu=0.01$, and $\lambda_j = 1/3$ for $j\in \{s,a,b\}$.}
\label{fig:tatonnement}
\end{figure}

\autoref{fig:tatonnement} illustrates the implications of \autoref{prop:transition_by_beta} and \autoref{prop:transition_by_alpha} in terms of allocation efficiency. The blue dashed line represents the allocation in the Boston mechanism equilibrium. When the mechanism changes to the DA, but students best respond to $r^B$, their rank-order choices become more homogeneous, especially when their unobservable preference types are near $1-v$. The following transitions, leading ultimately to the DA equilibrium, drawn in red solid line, represent the reinforcing cycle between more homogeneous rank-order choices and less information acquisition. In the sequence of students' behavior adjustments, the allocations become increasingly inefficient.

Finally, the self-reinforcing relationship between uniform rank-order reports and decreased information acquisition is intensified by the augmented expenses associated with obtaining information.
\begin{proposition}
\label{lem:intensifying_loop}
If $\mu_1 < \mu_2 < \lambda_s(1-v)$, then  $\overline{m}^H(r, \mu_1) < \overline{m}^H(r, \mu_2)$, and $g^H(\cdot ; r, \mu_1)$ is single crossing $g^H(\cdot; r, \mu_2)$ from below.\footnote{Similar to \autoref{prop:transition_by_alpha}, we assume $\mu_i < \lambda_s(1-v)$, $i\in \{1, 2\}$, to ensure that an interior best response under the DA mechanism exists for any $r \in (\hat{r}, 1)$.}
\end{proposition}

\subsection{Comparative statics in $\mu$.}

To compare the efficiency loss of the DA mechanism to the Boston mechanism, we initially used first-order dominance criteria to analyze the preference shock distributions of students matched with school $s$. However, this approach is unsuitable for comparative statics analysis. To overcome this challenge, we could employ a real-valued efficiency measure such as the average efficiency $W^{\Gamma} \equiv \frac{\int_{0}^1 \theta g^{\Gamma}(\theta) d\theta}{\lambda_s}$ across different mechanisms.\footnote{We divide by $\lambda_s$ because the mass of students who match with school $s$ is $\lambda_s$.} Nevertheless, calculating this average efficiency is challenging.

Alternatively, we can observe that higher levels of homogeneity in rank-order reports ($r^D, r^B > \hat{r}$) lead to the DA mechanism's efficiency loss, discouraging students from acquiring information at a cost. Therefore, for a comparative statics analysis, we will consider the real-valued measure of homogeneity.

\begin{proposition}
\label{prop:monotone_da}
Assume that $v > 1/2$ and $\lambda_s \leq \lambda_a$, which implies that school $s$ is always more competitive than school $a$ in any mechanism and for any $\mu \geq 0$.\footnote{This can be shown by observing that school $s$ is more competitive both in expected match payoffs, i.e., $E[u_s] > v + \frac{1}{2} > u_a=1$, and in ex-post preferences, i.e., $P[u_s > u_a] = v$, relative to the schools' capacities $v > \hat{r}=\frac{\lambda_s}{\lambda_s + \lambda_a}$.}
\begin{enumerate}
\item There exists $\overline{\mu} > 0$ such that, as $\mu$ increases from $0$ to $\overline{\mu}$, the equilibrium fraction $r^D$ under the DA mechanism increases from $v$ to $1$.
\item If $r^{B}_\mu$ and $r^{B}_{\mu'}$ are the equilibrium fractions of the Boston mechanism for $\mu$ and $\mu'$ with $\mu < \mu'$, respectively, and $r^{B}_{\mu} > \frac{1}{2} (\geq \hat{r})$, then $r^{B}_{\mu} < r^{B}_{\mu'}$. In addition, if $v \leq \frac{1}{2} + \frac{\lambda_b}{\lambda_s}$, then the equilibrium fraction $r^B_\mu$ of the Boston mechanism for any $\mu$ is bounded above by $\max\{\frac{1}{2}, 1-\lambda_a\}$.
\end{enumerate}
\end{proposition}
As $\mu$ increases, the equilibrium proportion of students who submit $sab$ under the Deferred Acceptance (DA) mechanism, denoted by $r^D$, increases and eventually converges to 1. In contrast, the proportion of students who submit $sab$ under the Boston mechanism, denoted by $r^B$, may also increase with $\mu$, but it remains relatively stable, provided that the capacity of a selective school $s$ is sufficiently smaller than the capacity of school $b$. More specifically, this occurs when $v \leq \frac{1}{2} + \frac{\lambda_b}{\lambda_s}$.

To demonstrate \autoref{prop:monotone_da} and its implications for the efficiency loss of the DA mechanism, we provide an illustrative numerical example. Consider a scenario with equal capacities ($\lambda_s = \lambda_a = \lambda_b = 1/3$) and two different values of $v$: $v = 0.6$ and $v = 0.7$. We then increase the value of $\mu$ from $0$, ensuring that an interior equilibrium exists in both the DA and Boston mechanisms.

\begin{figure}[h!]
 \centering
 \begin{subfigure}[t]{0.45\textwidth}
 \centering
 \includegraphics[scale=0.55]{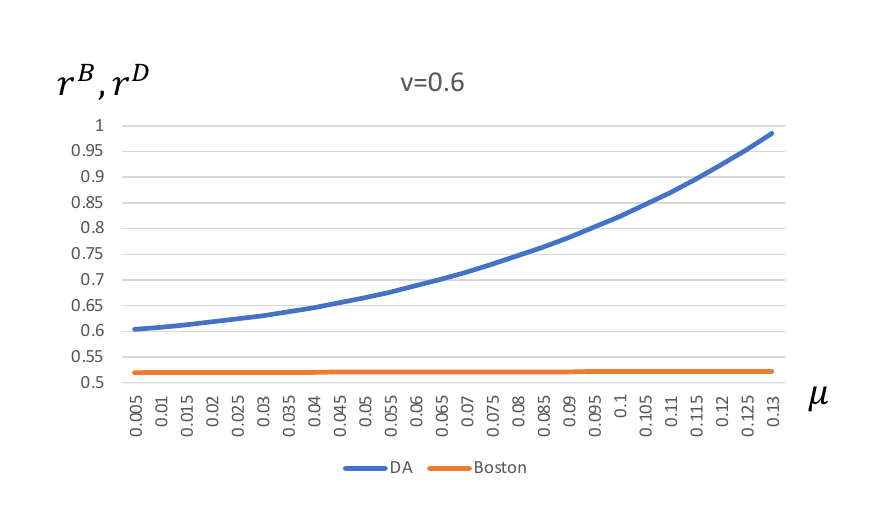}
 \caption{$v = 0.6$}
 \end{subfigure}
 ~
 \qquad
 \begin{subfigure}[t]{0.45\textwidth}
 \centering
 \includegraphics[scale=0.55]{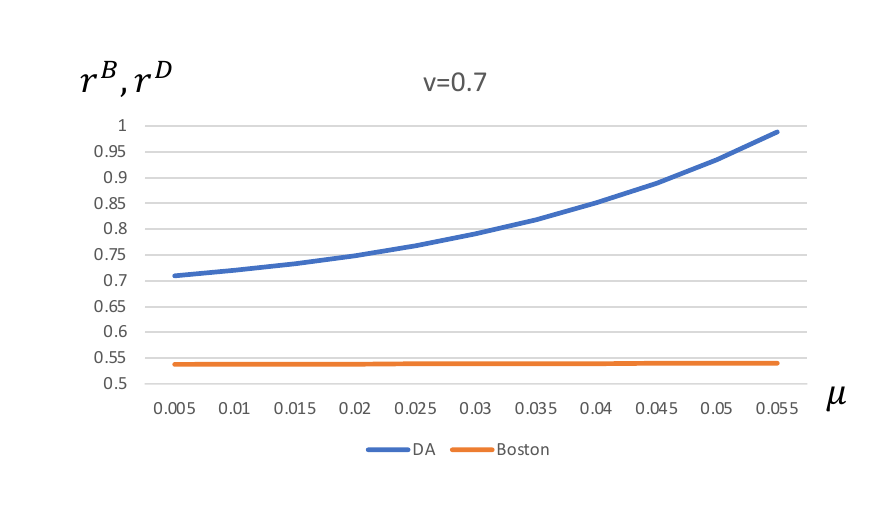}
 \caption{$v = 0.7$}
 \end{subfigure}
 \begin{subfigure}[t]{1\textwidth}
 \centering
 \includegraphics[scale=0.55]{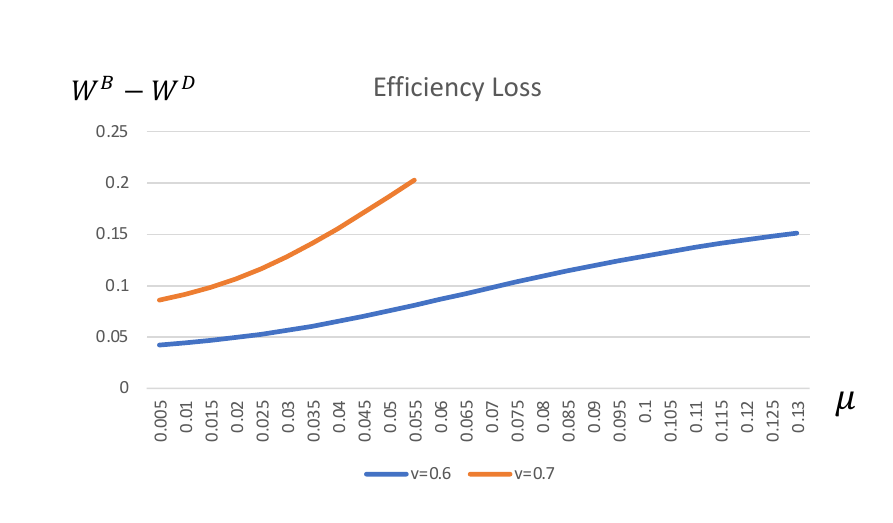}
 \caption{The DA mechanism's efficiency loss.}
 \end{subfigure}
  \caption{As $\mu$ increases, the homogeneity of rank-order reports to the DA mechanism increases rapidly, and results in an increasing efficiency loss.}
 \label{fig:eq_fraction_simulation}
\end{figure}

In \autoref{fig:eq_fraction_simulation}, Panels (a) and (b) illustrate that the proportion of students who report $sab$ in the DA mechanism rapidly approaches 1 as the cost of information acquisition, $\mu$, increases. However, the proportion of students reporting $sab$ in the Boston mechanism remains low and relatively stable. This finding suggests a potential for significant efficiency losses under the DA mechanism, particularly as the cost of information acquisition increases. Indeed, Panel (c) of \autoref{fig:eq_fraction_simulation} shows that the difference in average efficiency $W^{B} - W^{D}$ becomes more substantial as $\mu$ increases.

\section{Conclusion}
We analyze a school choice problem where students gather information on their preferences before submitting rank-order lists to a mechanism. We compare two widely used mechanisms: DA and Boston. The DA mechanism results in more uniform rank-order submissions, causing it to rely more on random tie-breaking than the Boston mechanism. As a result, students have less motivation to collect costly information, leading to further homogeneity in their rank-order submissions and greater efficiency loss. These characteristics, which result in the DA mechanism's efficiency loss, are amplified when the cost of information acquisition rises.

\appendix
\section{Appendix}

\subsection{Proof of \autoref{lem:eq_condition_da}}
For a given mechanism $\Gamma \in \{\text{Boston}, \text{DA}\}$, the equilibrium consistency $r^\Gamma = \int_0^1 m^{\Gamma}(\theta) d\theta$, where $m^{\Gamma}$ satisfies \eqref{eqn:m_uniquely_by_r}, implies that $r^{\Gamma}$ is a solution to the following equation:
$$
r = \int_0^1 \left(1+\frac{1-r}{r} \exp\left(-\frac{\Delta^{\Gamma} (\theta; r)}{\mu}\right)\right)^{-1} d\theta.
$$
Since $\Delta^{\Gamma}(\theta ; r) = \Delta^{\Gamma}(0; r) - \frac{\lambda_s \theta}{r}$ (as shown in \eqref{eqn:delta_boston} and \eqref{eqn:delta_da}), we can rewrite the above equation as follows:
\begin{eqnarray*}
r & = &\int_0^1 \left(1+\frac{1-r}{r}\exp\left(-\frac{\Delta^{\Gamma}(0; r)}{\mu}\right)\exp\left(\frac{\lambda_s \theta}{r\mu}\right)\right)^{-1} d\theta\\
& =& \frac{r \mu}{\lambda_s} \left.\log \left[
\frac{1-r}{r}\exp\left(-\frac{\Delta^{\Gamma}(0; r)}{\mu}\right)+ \exp\left(\frac{\lambda_s\theta}{r \mu}\right)
\right]\right|_0^1. 
\end{eqnarray*}
By canceling out $r$ and taking the exponential of each side, we obtain \eqref{eqn:eq_restriction_da}.

\subsection{Proof of \autoref{prop:equilibrium}}
\subsubsection{Proof for the DA mechanism}
\label{proof:da_equilibrium}

Given any $\mu > 0$, we can determine the upper and lower bounds $\overline{v}$ and $\underline{v}$, respectively, such that $v \in (\underline{v}, \overline{v})$ if and only if there exists a solution $r$ to the equilibrium condition \eqref{eqn:eq_restriction_da} under the DA mechanism in the interval $(\hat{r}, 1)$.

First, we find the upper bound $\overline{v}$. Let $z \equiv \frac{\lambda_s}{\mu}$, $w \equiv \frac{\lambda_a}{\mu}$, and $x \equiv \frac{1}{r} \in [1, \infty)$. Then, the equilibrium condition \eqref{eqn:eq_restriction_da} becomes:
\begin{align*}
e^z = 1 + \frac{e^{zx}-1}{
(x-1)e^{zx(1-v)}+1} \iff f(x, v) \equiv \frac{e^{zx}-1}{e^z-1}- (x-1)e^{zx(1-v)} =1. 
\end{align*}
For any $v$, we have $f(1 ; v)=1$ and $\lim_{x \to \infty} f(x , v) = \infty$. Moreover, we can show that:
\begin{align}
\frac{\partial f(x, v)}{\partial x} < 0 \iff & \frac{ze^{zx}}{e^z-1} - e^{zx(1-v)} - (x-1)e^{zx(1-v)} z(1-v) < 0 \notag\\
\iff & g(x, v) \equiv \frac{ze^{zxv}}{e^z-1} -1- (x-1)z(1-v)< 0.
\label{eqn:derivative_at_1_da}
\end{align}
Let $\overline{v}$ be such that:
\begin{equation}
g(1, \overline{v}) =0 \iff z e^{z\overline{v}} = e^z-1 \iff \overline{v} = \frac{1}{z}\log\left(\frac{e^z -1}{z}\right).\label{eqn:v_upper_bound}
\end{equation}
We can verify that $\overline{v} \in (1/2, 1)$ and is strictly increasing in $z$. We will use these properties later.

If $v < \overline{v}$, then $g(1 , v) < 0$ because the function $g$ is strictly increasing in $v$ and $g(1 , \overline{v}) = 0$. Moreover, $g(x, v)$ is strictly convex in $x$, and $\lim_{x \to \infty} g(x , v) = \infty$. Thus, if $v < \overline{v}$, then as $x$ increases from 1 to $\infty$, the function $g(x , v)$ changes sign exactly once from strictly negative to strictly positive. This implies that $f'(x, v) < 0$ at $x=1$ and $f(x, v)$ is a strictly single-dipped function of $x$. Thus, a unique solution $x^* > 1$ of $f(x;v)=1$ exists (hence, a unique solution $r^* = \frac{1}{x^*}$ of \eqref{eqn:eq_restriction_da} exists in the interval $(0,1)$).

In contrast, if $v \geq \overline{v}$, then $g(1 ; v) \geq 0$. Moreover, for any $x >1$, we have
\begin{equation}
\label{eqn:finding_underlinev_overlinev}
\frac{\partial g(x , v)}{\partial x} > \frac{\partial g(1 , v)}{\partial x} \geq (z\overline{v}) \frac{ze^{z\overline{v}}}{e^z-1} - z(1-\overline{v}) \geq z (2\overline{v} - 1) \geq 0,
\end{equation}
where the last two inequalities follow from \eqref{eqn:v_upper_bound} and the fact that $\overline{v} \geq 1/2$, respectively. Thus, if $v \geq \overline{v}$, then $g(x ; v) > 0$ for every $x >1$, and there exists no solution $x >1$ to $f(x;v)=1$.

Second, we find the lower bound $\overline{v}$. Let $\hat{x} \equiv \frac{1}{\hat{r}} = \frac{\lambda_s + \lambda_a}{\lambda_s} >1$ and define $\underline{v}$ such that \begin{align*}
f(\hat{x}; \underline{v})=1 & \iff \frac{e^{z+w}-1}{e^{z}-1} - \frac{w}{z}e^{(z+w)(1-\underline{v})} = 1 
\iff e^{(z+w)\underline{v}} = \frac{w}{1 - e^{-w}} \frac{e^z - 1}{z}\\
& \iff \underline{v} = \frac{1}{z+w}\log\left(\frac{w}{1 - e^{-w}} \frac{e^z-1}{z}\right).
\end{align*}
We previously defined $\overline{v}$ such that $f(1; \overline{v}) = 1$. Moreover, $\frac{\partial f(x, \overline{v})}{\partial x} > 0$ for every $x > 1$ because  \eqref{eqn:derivative_at_1_da} holds for the opposite inequalities, and $g(x, \overline{v}) > 0$ for every $x > 1$, as we observed after \eqref{eqn:finding_underlinev_overlinev}. Thus, $f(\hat{x}, \overline{v}) > 1 = f(\hat{x},\underline{v})$, which implies $\overline{v} > \underline{v}$.

We previously proved that $v < \overline{v}$ if and only if a unique solution $x^*$ of $f(x, v) =1$ is greater than $1$ (i.e., a unique equilibrium proportion $r^* < 1$). The function $f(x, v)$ is strictly and continuously increasing in $v$ for every $x>1$. Therefore, the unique solution $x^*$ is in $(1, \hat{x})$ (i.e., $r^* \in (\hat{r}, 1)$) if and only if $v \in (\underline{v}, \overline{v})$.

% Last, $$\lim_{\mu \to 0} \overline{v} = \lim_{z \to \infty} \frac{1}{z} \log\left(\frac{e^z - 1}{z}\right) = 1,\quad \text{and} \quad \lim_{\mu \to \infty} \overline{v} = \lim_{z \to 0} \frac{1}{z} \log\left(\frac{e^z - 1}{z}\right) = \frac{1}{2}.$$ Moreover,\begin{align*}\underline{v} & = \frac{1}{z+w}\left[\log\left(\frac{e^z-1}{z}\right) + w - \log\left(\frac{e^w-1}{w}\right)\right]\\& = \hat{r} \left[\frac{1}{z} \log\left(\frac{e^z-1}{z}\right)\right] +(1-\hat{r}) \left[ 1 - \frac{1}{w}\log\left(\frac{e^w-1}{w}\right)\right], \end{align*}implies $$ \lim_{\mu \to 0} \underline{v} = \lim_{z, w \to \infty} \underline{v} = \hat{r}, \quad \text{and} \quad \lim_{\mu \to \infty} \underline{v} = \lim_{z, w \to 0} \underline{v} =\frac{1}{2}.$$

\subsubsection{Proof for the Boston mechanism}

Suppose that the DA mechanism has an interior equilibrium with $r^D \in (\hat{r}, 1)$, i.e., $v \in (\underline{v}, \overline{v})$. Then, $r^D$ is the unique solution of \eqref{eqn:eq_restriction_da} for the DA mechanism in the interval $(\hat{r}, 1)$.

Note that $r > \hat{r}$ if and only if 
$$\Delta^{D}(0; r) - \Delta^{B}(0; r) =\min\left\{\frac{\lambda_a}{1-r}, \frac{1-\lambda_a}{r} \right\}- \frac{\lambda_s}{r} > 0$$
When $r < \hat{r}$ (or, $r = \hat{r}$), the opposite strict inequality (or, the equality) holds. Therefore, the right-hand side of \eqref{eqn:eq_restriction_da} under the Boston mechanism is smaller than, larger than, or equal to the right-hand side under the DA mechanism, when $r > \hat{r}$, $r < \hat{r}$, or $r = \hat{r}$, respectively. As a result, an interior equilibrium of the Boston mechanism exists with $r^{B} \in (\hat{r}, r^D)$ by the Intermediate Value Theorem.

We next prove that the equilibrium proportion $r^B \in (\hat{r}, r^D)$ under the Boston mechanism is unique.

Let $z \equiv \frac{\lambda_s}{\mu}$, $w \equiv \frac{\lambda_a}{\mu}$, and $y \equiv \frac{\lambda_b}{\mu}$. The equilibrium condition \eqref{eqn:eq_restriction_da} for the Boston mechanism becomes
\begin{equation}
\label{eqn:boston_eq_condition_new}
e^z = 1 + \frac{e^{\frac{z}{r}} -1}{\frac{1-r}{r} e^{\min\{\alpha_1(r), \alpha_2(r)\}} e^{-\frac{zv}{r}} +1} \equiv h(r),
\end{equation}
where $\alpha_1(r) \equiv \frac{w}{1-r}$ and $\alpha_2(r) \equiv \frac{y+z}{r}$.

We search for the solution to \eqref{eqn:boston_eq_condition_new} in two regions. Define $h_i(r) \equiv 1 + \frac{e^{\frac{z}{r}} -1}{\frac{1-r}{r} e^{\alpha_i(r)} e^{-\frac{zv}{r}} +1}$ for $i=1,2$. Then, $h(r) = \max\{h_1(r), h_2(r)\}$. More precisely, if $r \leq 1-\lambda_a$, then $\alpha_1(r) \leq \alpha_2(r)$, so $h(r) = h_1(r)$; if $r \geq 1-\lambda_a$, then $\alpha_1(r) \geq \alpha_2(r)$, so $h(r) = h_2(r)$. We make two observations below regarding $h_1(r)$ and $h_2(r)$.

\begin{claim}
\label{claim:single_crossing_boston_rhs}
$h_1(r)$ is single crossing $e^z$ from above to below as $r$ increases in $[\hat{r}, 1)$. Specifically, if $e^z \geq h_1(r')$ for some $r' \in [\hat{r}, 1)$, then $e^z > h_1(r'')$ for every $r'' \in (r', 1)$.
\end{claim}

\begin{proof}
For any $r \in [\hat{r}, 1)$,
\begin{align*}
e^z \geq h_1(r) & \iff \frac{1-r}{r} e^{\alpha_1(r)}e^{-\frac{zv}{r}} +1 \geq \frac{e^{\frac{z}{r}} -1}{e^z -1}\\& \iff e^{\frac{w}{1-r}}e^{-\frac{zv}{r}} \geq \frac{e^z}{e^z -1}\frac{(e^{z((1/r)-1)} -1)}{(1/r)-1}.
\end{align*}
Moreover, the strict inequalities can be used instead of the weak ones. The left-hand side of the last inequality is strictly increasing in $r$, whereas the right-hand side is strictly decreasing in $r$ because
$$
\left(\frac{e^{z(x-1)} -1}{x-1}\right)' > 0 \iff z(x-1)e^{z(x-1)} > e^{z(x-1)}-1 \iff e^{-z(x-1)} > 1-z(x-1),
$$
which holds for every $x > 1$. Therefore, \autoref{claim:single_crossing_boston_rhs} holds.
\end{proof}

\begin{claim}
$h_2(r)$ is single crossing $e^z$ from above to below as $r$ increases in $[\hat{r}, 1)$. Formally, if $e^z \geq h_2(r')$ for some $r' \in [\hat{r}, 1)$, then $e^z > h_2(r'')$ for every $r'' \in (r', 1)$.
\label{claim:single_dipped_boston_rhs}
\end{claim}

\begin{proof}
For any $r \in [\hat{r}, 1)$, 
$$
e^z \geq h_2(r)  \iff 
1 \geq \frac{e^{\frac{z}{r}} -1}{e^z -1}- \frac{1-r}{r} 
e^{\alpha_2(r)}e^{-\frac{zv}{r}}.
$$
We make a change of variable $x \equiv \frac{1}{r} \in (1, 1/\hat{r}]$ and write $\tilde{\alpha}_2(x) = (y+z)x$, as defined in \eqref{eqn:boston_eq_condition_new}. Then, 
\begin{equation}
e^z \geq h_2(r) \iff 1 \geq \frac{e^{zx} -1}{e^z -1}- (x-1) 
e^{\tilde{\alpha}_2(x)}e^{-zvx} \equiv \tilde{h}_2(x).\label{eqn:boston_step}
\end{equation}
Note that $\tilde{\alpha}'_2 = y+z$ is independent of $x$. Hence,
\begin{align*}
\tilde{h}_2'(x) < 0 & \iff 0 > \frac{ze^{zx}}{e^z-1} - e^{\tilde{\alpha}_2(x) - zvx} - (x-1)(\tilde{\alpha}'_2 - zv)e^{\tilde{\alpha}_2(x) - zvx}\\& \iff 0 > \frac{ze^{z(1+v)x - \tilde{\alpha}_2(x)}}{e^z-1} -1- (x-1)(\tilde{\alpha}'_2 - zv) \equiv \tilde{g}(x, v).
\end{align*}

First, $\tilde{h}_2(1) = 1$.

Second, we note that $v \in (\underline{v}, \overline{v})$, i.e., an interior equilibrium of the DA mechanism where $r^D \in (\hat{r}, 1)$ exists uniquely. We also note that $g(x, v)$, defined in \eqref{eqn:derivative_at_1_da} for the DA mechanism, satisfies $g(1, v) <0$. Since $\tilde{g}(1, v) < g(1; v)$, we have $\tilde{g}(1, v) < 0$, which implies that $\tilde{h}_2'(1) < 0$.

Last, if $zv > y$, then $\tilde{g}(x,v)$ is a strictly convex function of $x \in (1, 1/\hat{r}]$. Consequently, as $x$ increases from $1$ to $\frac{1}{\hat{r}}$, either $\tilde{g}(x, v)$ remains to be strictly negative, or its sign changes exactly once from strictly negative to strictly positive. On the other hand, if $zv \leq y$, then $\tilde{g}(x, z)$ is decreasing in $x$, so it remains to be strictly negative. In either case, $\tilde{h}_2(x)$ is a strictly single-dipped function of $x \in (1, 1/\hat{r}]$.

Altogether, since $\tilde{h}_2(1)=0$, $\tilde{h}_2'(1) < 0$, and $\tilde{h}_2(x)$ is a strictly single-dipped function of $x \in (1, 1/\hat{r}]$, the function $\tilde{h}_2(x)$ single crosses $1$ from below to above as $x$ increases in $(1, 1/\hat{r}]$. Consequently, from \eqref{eqn:boston_step}, we infer that $h_2(r)$ single crosses $e^z$ from above to below as $r$ increases in $[\hat{r}, 1)$.
\end{proof}

The uniqueness of the interior equilibrium fraction $r^B \in (\hat{r}, r^D)$ under the Boston mechanism follows from the following reasoning. If $h(1-\lambda_a) \geq e^z$, then there is no solution of $h(r) = e^z$ in $(\hat{r}, 1-\lambda_a)$ according to \autoref{claim:single_crossing_boston_rhs}, and there is at most one solution in $[1-\lambda_a, 1)$ according to \autoref{claim:single_dipped_boston_rhs}. On the other hand, if $h(1-\lambda_a) < e^z$, then there is at most one solution of $h(r) = e^z$ in $(\hat{r}, 1-\lambda_a]$ according to \autoref{claim:single_crossing_boston_rhs}, and no solution exists in $[1-\lambda_a, 1)$ according to \autoref{claim:single_dipped_boston_rhs}.

\subsection{Proof of \autoref{prop:boston_efficient_info}}

For any mechanism $\Gamma \in \{\text{Boston}, \text{DA}\}$, we use equations \eqref{eqn:delta_boston} and \eqref{eqn:delta_da} to apply $\Delta^{\Gamma}(\theta ; r) = \Delta^{\Gamma}(0; r) - \frac{\lambda_s \theta}{r}$ to the equilibrium condition \eqref{eqn:m_uniquely_by_r}. This yields:
$$
m^{\Gamma}(\theta) = \left( 1 + \frac{1-r^{\Gamma}}{r^{\Gamma}} \exp\left(-\frac{\Delta^{\Gamma}(0 ; r^{\Gamma})}{\mu}\right) \exp\left(\frac{\lambda_s \theta}{r\mu}\right)\right)^{-1}.
$$
Another equilibrium condition \eqref{eqn:eq_restriction_da} implies that:
$$
\frac{1-r^{\Gamma}}{r^{\Gamma}} \exp\left(-\frac{\Delta^{\Gamma}(0 ; r^{\Gamma})}{\mu}\right) = \frac{\exp\left(\frac{\lambda_s}{r^{\Gamma} \mu}\right) -1}{\exp \left(\frac{\lambda_s}{\mu}\right) - 1}-1.
$$
Thus, if we let $z \equiv \frac{\lambda_s}{\mu}$, we have:
$$
m^{\Gamma}(\theta) = \left[ 1 + h(\theta, r^\Gamma)\right]^{-1}, \quad \text{where}\quad 
h(\theta, r) = \frac{\exp\left(- \frac{z\theta}{r}\right)(\exp\left(\frac{z}{r}\right) - \exp(z))}{\exp(z) - 1}.
$$
Since $g^{\Gamma}(\theta) = m^{\Gamma}(\theta) \frac{\lambda_s}{r^{\Gamma}}$, we get:
$$
\frac{d (\lambda_s / g^{\Gamma}(\theta))}{d \theta} = \frac{\partial (r^{\Gamma} (1 + h(\theta, r^{\Gamma}))}{\partial \theta} = - z \cdot h(\theta, r^{\Gamma}).
$$
Moreover, 
\begin{align*}
\frac{\partial h(\theta, r)}{\partial r} 
& = \left[\frac{z\theta}{r^2}e^{- \frac{z\theta}{r}} (e^{\frac{z}{r}} - e^z) - \frac{z}{r^2} e^{-\frac{z\theta}{r}} e^{\frac{z}{r}}\right] \frac{1}{e^z-1}\notag\\
&= - \left[\frac{z}{r^2}e^{- \frac{z\theta}{r}} (e^{\frac{z}{r}}(1-\theta) + e^z \theta)\right] \frac{1}{e^z-1} < 0.
\end{align*}
Therefore, 
\begin{align*}
r^B < r^D \implies 
& (\forall \theta \in (0,1)) \quad h(\theta, r^B)> h(\theta, r^D)\\
\implies & (\forall \theta \in (0,1)) \quad \frac{d (\lambda_s / g^{B}(\theta))}{d \theta} < \frac{d (\lambda_s / g^{D}(\theta))}{d \theta}\\
\implies & (\forall \theta < \theta') \quad \frac{1}{g^{B}(\theta)} \leq \frac{1}{g^{D}(\theta)} \implies \frac{1}{g^{B}(\theta')} < \frac{1}{g^{D}(\theta')}\\
\implies & \text{$g^{B}(\theta)$ is single crossing $g^{D}(\theta)$ from below.} 
\end{align*}

\subsection{Proof of \autoref{prop:transition_by_beta}}

Expressions \eqref{eqn:delta_boston} and \eqref{eqn:delta_da} show that both $\frac{\Delta^{D}(\theta; r^B)}{\mu}$ and $\frac{\Delta^{B}(\theta; r^B)}{\mu}$ are functions of $\theta$ in the form of $\alpha(\theta + \beta)$, where $\alpha = \frac{\lambda_s}{r^B \mu}$ and $\beta = v-1$ for the Deferred Acceptance (DA) mechanism, while for the Boston mechanism, $\beta$ is smaller than $v-1$. Therefore, switching the mechanism from Boston to DA, while keeping $r^B$ unchanged, corresponds to increasing $\beta$ from $\underline{\beta} = v - \min\left\{\frac{r^B}{\lambda_s} \frac{\lambda_a}{1-r^B}, \frac{ 1-\lambda_a}{\lambda_s}\right\}$ to $\overline{\beta} = v-1$.

According to \cite{yang2015coordination}, a student's interior optimal strategy $m:[0,1] \to [0,1]$, for a given $\alpha$ and $\beta$, satisfies the following first-order condition:
$$
\alpha (\theta + \beta) = \ln \left(\frac{m(\theta)}{1-m(\theta)}\right) - \ln \left(\frac{\overline{m}}{1-\overline{m}}\right).
$$
where $\overline{m}$ denotes the expected value of $m(\theta)$ over the uniform distribution of $\theta$. Thus, an optimal strategy has the form $m(\theta)= \frac{L e^{\alpha(\theta + \beta)}}{L e^{\alpha(\theta + \beta)} + 1}$, where $L = \frac{\overline{m}}{1-\overline{m}}$ is the likelihood ratio between the reports $sab$ and $asb$. The consistency $\overline{m} = \int m(\theta) d\theta$ requires the likelihood ratio $L$ to satisfy
\begin{equation}
\label{eqn:consistency}
\int_0^1 \frac{L e^{\alpha(\theta + \beta)}}{L e^{\alpha(\theta + \beta)} + 1} d\theta = \frac{L}{L+1}\iff  \int_0^1 \frac{1}{L+e^{-\alpha(\theta + \beta)}} d\theta = \frac{1}{L+1}.
\end{equation}

\bigskip

To begin, we establish the existence of a unique interior solution $L$ to equation \eqref{eqn:consistency}. Finding a solution $L$ to \eqref{eqn:consistency} is equivalent to finding a solution $\overline{m} = \frac{L}{L+1} \in (0,1)$ to the equation:
\begin{align}
& \int_0^1 \frac{\overline{m} e^{\alpha(\theta +\beta)}}{\overline{m} e^{\alpha(\theta +\beta)} + 1-\overline{m}} d\theta  -  \overline{m} = 0\notag\\
& \iff \left.\log (\overline{m} e^{\alpha(\theta + \beta)} + 1-\overline{m})\right|^1_0 - \alpha \overline{m} = 0\notag\\
& \iff f(\overline{m}; \alpha, \beta) \equiv \overline{m} (e^{\alpha(1+\beta)} - e^{\alpha (\overline{m} + \beta)}) + (1-\overline{m})(1 - e^{\alpha \overline{m}})=0.\label{eqn:solve_for_mbar}
\end{align}
Since $v \in (\underline{v}, \overline{v})$, the Boston mechanism (as well as the DA mechanism) has a unique interior equilibrium with the proportion $r^B \in (\hat{r}, 1)$ of students reporting $sab$. Therefore, we have $f(r^B ; \alpha, \underline{\beta}) = 0$.

For any $\beta^* \in (\underline{\beta}, \overline{\beta}]$, we have $0 = f(r^B ; \alpha, \underline{\beta}) < f(r^B ; \alpha, \beta^)$ because the function $f$ is increasing in $\beta$. Furthermore, $f(1; \alpha, \beta^) = 0$, and since $1+\beta^* \leq 1+\overline{\beta} = v < \overline{v}$, we can see that
$$f'(1; \alpha, \beta^) = e^{\alpha} -1 - \alpha e^{\alpha(1+\beta^)} > e^z -1 - ze^{z\overline{v}} = 0 \quad \text{(by \eqref{eqn:v_upper_bound})}.$$
Therefore, there exists $\overline{m} \in (r^B, 1)$ such that $f(\overline{m} ; \alpha, \beta^) = 0$, which is equivalent to $L^*= \frac{\overline{m}}{1-\overline{m}}$ being a solution of \eqref{eqn:consistency} at $\beta^* > \underline{\beta}$.

\bigskip
The solution $L^*$ of \eqref{eqn:consistency} at $\beta^* > \underline{\beta}$ must be unique because
$$\frac{d}{dL}\left(\int_0^1 \frac{1}{e^{-\alpha(\theta + \beta^*)} + L^*} d\theta - \frac{1}{L^*+1}\right) = - \int_0^1 \frac{1}{(e^{-\alpha(\theta+\beta^*)} + L^*)^2} d\theta + \frac{1}{(L^* + 1)^2} < 0,
$$
The last inequality holds because $\frac{1}{e^{-\alpha(\theta + \beta^*)} + L^*}$ is a random variable whose expected value is $\frac{1}{L^*+1}$ by \eqref{eqn:consistency}, and for a random variable $X$, $Var[X] = E[X^2] - (E[X])^2 > 0$.

\bigskip
Next, $L(\beta)$ be the unique solution to \eqref{eqn:consistency} for each $\beta \in (\underline{\beta}, \overline{\beta})$. By the Implicit Function Theorem, we have
$$
\int_0^1 \frac{-\alpha e^{-\alpha(\theta + \beta)} + L'(\beta)}{(e^{-\alpha(\theta + \beta)} + L(\beta))^2} d\theta = \frac{L'(\beta)}{(L(\beta)+1)^2}, 
$$
which implies
$$
L'(\beta) \left[\int_0^1 \frac{1}{(e^{-\alpha(\theta+\beta)} + L(\beta))^2} d\theta - \frac{1}{(L(\beta) + 1)^2}\right] = \int_0^1 \frac{\alpha  e^{-\alpha(\theta + \beta)}}{(e^{-\alpha(\theta + \beta)} + L(\beta))^2} d\theta > 0. 
$$
Therefore, $L'(\beta) > 0$.

\bigskip
Finally, take any $\beta_1 < \beta_2$ in $(\underline{\beta}, \overline{\beta})$, and let $L_1$ and $L_2$ be the solutions to \eqref{eqn:consistency} for $\beta_1$ and $\beta_2$, respectively. Then, $L_1 < L_2$, as we showed above. For each $\beta_i$, $i \in \{1,2\}$, let  $m_i(\theta)$ be the optimal strategy and $g_i(\theta) \equiv m_i(\theta)\frac{\lambda_s}{\overline{m}_i}$ be the resulting assignment of students to school $s$ by the DA mechanism. Then, $\frac{\lambda_s}{g_i(\theta)} = \frac{L + e^{-\alpha(\theta + \beta_i)}}{L_i+1}$, and
\begin{align*}
\frac{\lambda_s}{g_2(\theta)} - \frac{\lambda_s}{g_1(\theta)} = \left(\frac{e^{-\alpha \beta_2}}{L_2+1} - \frac{e^{-\alpha \beta_1}}{L_1+1} \right) e^{-\alpha \theta} + \left(\frac{L_2}{L_2+1} - \frac{L_1}{L_1+1} \right),
\end{align*}
which is strictly increasing in $\theta$. Hence, $g_1$ is single crossing $g_2$ from below. 

\subsection{Proof of Propositions  \ref{prop:transition_by_alpha} and \ref{lem:intensifying_loop}}

Propositions \ref{prop:transition_by_alpha} and \ref{lem:intensifying_loop} both analyze the student's problem under the DA mechanism, but with different assumptions about the cost of information $\mu$ or the beliefs of the other students, particularly the proportion $r$ of other students reporting $sab$.

For any $\mu$ and $r$ such that $\mu < \lambda_s(1-v)$ and $r \in (r_\mu^B, r_{\mu}^D)$, we have $\frac{\Delta^{D}(\theta; r)}{\mu} = \alpha (\theta + v-1)$, where $\alpha = \frac{\lambda_s}{r \mu}$. A decrease of $\alpha$ corresponds to an increase of $r$ when $\mu$ is fixed (for \autoref{prop:transition_by_alpha}) or an increase of $\mu$ when $r$ is fixed (for \autoref{lem:intensifying_loop}).

In the previous proof of \autoref{prop:transition_by_beta}, we showed that a student's interior optimal strategy is $m(\theta)=\frac{\overline{m}e^{\alpha(\theta + \beta)}}{\overline{m} e^{\alpha(\theta + \beta)} + 1-\overline{m}}$ where $\overline{m}$ is a unique interior solution of \eqref{eqn:solve_for_mbar}, or equivalently
$$
h(\overline{m}, \alpha) \equiv e^{-\alpha \overline{m}} f(\overline{m};\alpha, v-1) = \overline{m} (e^{-\alpha (\overline{m}-v)} - e^{-\alpha (1-v)}) + (1-\overline{m})(e^{-\alpha \overline{m}}-1)=0.
$$
Note that we vary $\alpha$ but hold $\beta$ fixed at $v-1$ because we are considering only the DA mechanism.

\bigskip
(\underline{Part 1})

First, we show that the unique solution $\overline{m}^*$ of $h(\overline{m}, \alpha) = 0$ in $(\hat{r}, 1)$ decreases in $\alpha$. This result implies that the homogeneity of rank-order reports increases in the agents' belief $r$ (\autoref{prop:transition_by_alpha}) and in the cost of information acquisition $\mu$ (\autoref{lem:intensifying_loop}).

Note that $h(0, \alpha) = h(1, \alpha) =0$. Moreover, 
\begin{eqnarray*}
&&\frac{\partial h(0 , \alpha)}{\partial \overline{m}} = e^{\alpha v} (1-e^{-\alpha}) -\alpha \geq e^{\alpha/2} - e^{-\alpha/2} - \alpha > 0, \quad \text{and}\\
&&\frac{\partial h(1, \alpha)}{\partial \overline{m}} = e^{-\alpha} (e^{\alpha} -1 - \alpha e^{\alpha v}) > e^{-\alpha} (e^{\alpha} -1 - \alpha e^{\alpha \overline{v}}) > 0,
\end{eqnarray*}
where the last inequality holds because $e^{\alpha} -1 - \alpha e^{\alpha \overline{v}}$ increases in $\alpha$, $e^z -1 - ze^{z\overline{v}} = 0$ by \eqref{eqn:v_upper_bound}, and $\alpha \equiv \frac{\lambda_s}{r \mu} > \frac{\lambda_s}{\mu}\equiv z$. Together, these inequalities imply that at a unique interior solution $\overline{m}^* \in (0,1)$ of $h(\overline{m}, \alpha) = 0$, it must be that $\frac{\partial h(\overline{m}^*, \alpha)}{\partial \overline{m}} < 0$.

On the other hand, since $\mu < \lambda_s (1-v)$, we have $\frac{1}{\alpha} + v < \frac{\mu}{\lambda_s} + v < 1$, which implies that, at $\overline{m} = \frac{1}{\alpha} + v (< 1)$,
\begin{align*}
& h(\overline{m}, \alpha) = \frac{1}{\alpha e} \left((1+\alpha v)(1-e^{1-\alpha(1-v)}) + (\alpha (1-v)-1)(e^{\alpha v} - e) \right) > 0,
\end{align*}
which implies $\overline{m}^* > \frac{1}{\alpha} + v$. Then, $1 < \alpha(\overline{m}^* -v) < \alpha(1-v)$, and 
\begin{align*}
\frac{\partial h(\overline{m}^*, \alpha)}{\partial \alpha} & <\overline{m}^*\frac{\partial }{\partial \alpha}\left(e^{-\alpha(\overline{m}^*-v)} - e^{-\alpha(1-v)}\right) \\
& =
\frac{1}{\alpha} \left(-\alpha(\overline{m}^*-v)e^{-\alpha (\overline{m}^*-v)} + \alpha(1-v) e^{-\alpha(1-v)}\right) < 0,
\end{align*}
where the last inequality holds because $xe^{-x}$ decreases in $x > 1$. 

Applying the Implicit Function Theorem, we can conclude that the solution $\overline{m}^*$ decreases in $\alpha$ (or, increases in $r$ and $\mu$). 
\bigskip

(\underline{Part 2})

Let $L(\alpha)$ denote the unique solution of \eqref{eqn:solve_for_mbar}. Then, we have $m(\theta; \alpha) = \frac{L(\alpha) e^{\alpha(\theta + v-1)}}{L(\alpha) e^{\alpha(\theta + v-1)} + 1}$ and
$$
\frac{g(\theta; \alpha)}{\lambda_s} = \frac{L(\alpha) + 1}{L(\alpha) + e^{-\alpha(\theta +v - 1)}}.
$$

In Part 1, we showed that $L(\alpha) = \frac{\overline{m}(\alpha)}{1-\overline{m}(\alpha)}$ decreases as $\alpha$ increases. Furthermore, we have
\begin{align*}
\frac{\partial g(\theta;\alpha)}{\partial \alpha} > 0
& \iff L'(\alpha) (L(\alpha) + e^{-\alpha(\theta +v - 1)}) > (L(\alpha)+1) (L'(\alpha) - (\theta + v - 1) e^{-\alpha(\theta +v - 1)})\\
&\iff \theta + v - 1 > \frac{L'(\alpha)}{L(\alpha)+1} (e^{\alpha(\theta + v - 1)} - 1). 
\end{align*}
Thus, we can conclude that $\frac{d g(\theta; \alpha)}{d\alpha} > 0$ for every $\theta > 1-v$, and $\frac{d g(\theta; \alpha)}{d\alpha} < 0$ for every $\theta < 1-v$. Therefore, for $\alpha_1 > \alpha_2$, $g^H(\cdot ; \alpha_1)$ is single crossing $g^H(\cdot ; \alpha_2)$ from below. 

\subsection{Proof of \autoref{prop:monotone_da}}

Given the parameter values $v > 1/2$ and $\lambda_s \leq \lambda_a$, we observe that $v > \frac{1}{2} \geq \hat{r} \equiv \frac{\lambda_s}{\lambda_s + \lambda_a}$.

\subsubsection{Proof for the DA mechanism}

For a given $\mu$, let $\overline{v}_{\mu}$ denote the upper bound of $v$ for which an interior equilibrium with a fraction $r^D_{\mu} \in (\hat{r}, 1)$ exists. Specifically, we have $\overline{v}_{\mu} = \frac{1}{z}\log\left(\frac{e^z -1}{z}\right)$ where $z \equiv \frac{\lambda_s}{\mu}$, as defined in \eqref{eqn:v_upper_bound}. It can be shown that $\overline{v}_{\mu}$ is a strictly increasing and continuous function of $z$, ranging from $\lim_{z \to 0} \overline{v}_{\mu} = \frac{1}{2}$ to $\lim_{z \to \infty} \overline{v}_{\mu} = 1$. Conversely, for each $v  > \frac{1}{2}$, we can find $\overline{\mu}$ such that $v = \frac{1}{\underline{z}}\log\left(\frac{e^{\underline{z}} -1}{\underline{z}}\right)$, where $\underline{z} \equiv \frac{\lambda_s}{\overline{\mu}}$. An interior equilibrium under the DA mechanism exists with a fraction $r^D_{\mu} \in (\hat{r}, 1)$ only if $\mu < \overline{\mu}$.
Additionally, $\lim_{\mu \to \overline{\mu}} \overline{v}_{\mu} = v$ (i.e., the upper bound for an interior equilibrium becomes close to $v$), and so $\lim_{\mu \to \overline{\mu}} r^D_{\mu} = 1$.

\bigskip
We will now demonstrate that the equilibrium fraction $r^D_\mu$ strictly increases as $\mu$ increases.

Let $x \equiv \frac{1}{r}$ and $z \equiv \frac{\lambda_s}{\mu}$. We can express the equilibrium condition \eqref{eqn:eq_restriction_da} as follows:

\begin{align*}
& e^z = 1 + \frac{e^{zx}-1}{
(x-1)e^{zx(1-v)}+1} \\
& \iff e^{zx}-1 - (e^z -1)((x-1)e^{zx(1-v)}+1)=0\\
& \iff (e^{zx} - e^z) - (x-1)(e^z-1)e^{zx(1-v)}=0\\
& \iff h(x; z) \equiv e^{zxv}(1-e^{z(1-x)}) - (x-1)(e^z-1)=0. 
\end{align*}
By \autoref{prop:equilibrium}, there exists a unique solution $x^*$ of $h(x ; z) = 0$. 

We claim the following in order to apply the Implicit Function Theorem. For ease of exposition, we will suppress the dependency of $x^*$ on $z$.

\begin{claim}
\label{claim:auxiliary}
\begin{enumerate}
\item $\frac{1}{y} - \frac{1}{e^y-1} < \frac{1}{2}$ for any $y > 0$, and
\item $\log\left(\frac{y}{e^y-1}\right)+ y + \frac{y}{e^y-1}$ is strictly increasing in $y$. 
\end{enumerate}
\end{claim}

\begin{proof}
(\underline{Part 1}) By L'Hopital's rule,
\begin{align*}
\lim_{y \to 0} \frac{1}{y} - \frac{1}{e^y-1}
& = \lim_{y \to 0} \frac{e^y-1-y}{y(e^y-1)}=\lim_{y\to 0} \frac{e^y-1}{e^y-1 + ye^y} = \lim_{y\to 0} \frac{e^y}{2e^y + ye^y}= \frac{1}{2}.
\end{align*}
Moreover, 
\begin{align}
&\left(\frac{1}{y} - \frac{1}{e^y-1}\right)' = - \frac{1}{y^2} +\frac{e^y}{(e^y-1)^2} < 0 \iff y^2 e^y < (e^y - 1)^2 \label{eqn:claim_intermediate_step}\\
& \impliedby 2ye^y + y^2 e^y < 2(e^y-1) e^y \iff 2y + y^2 < 2(e^y-1) \notag\\
& \impliedby 2 + 2y < 2e^y,\notag
\end{align}
which holds for every $y > 0$. In each step above, we observed that the two sides of an inequality converge to each other as $y \to 0$, and we compared their derivatives.

\bigskip

(\underline{Part 2}) 

Let $f(y) \equiv \log\left(\frac{y}{e^y-1}\right)+ y + \frac{y}{e^y-1}$. Then,
\begin{align*}
& f'(y) = 1 + \left(1 + \frac{e^y -1}{y}\right) \left(\frac{y}{e^y-1}\right)' = 1 + \left(1 + \frac{e^y -1}{y}\right) \frac{e^y-1 - ye^y}{(e^y-1)^2} > 0\\
& \iff \left(1 + \frac{e^y-1}{y}\right)\left(1 - \frac{ye^y}{e^y-1}\right) > 1-e^y\\
& \iff \frac{e^y-1}{y} > \frac{ye^y}{e^y-1},
\end{align*}
which always holds by \eqref{eqn:claim_intermediate_step}.
\end{proof}

\begin{claim}
$\frac{\partial h (x^*, z)}{\partial x} >0$.
\label{claim:derivative_h_x_positive}
\end{claim}

\begin{proof}
Observe that
\begin{align*}
\frac{\partial h (x, z)}{\partial x} 
& = e^{zxv} (zv)(1-e^{z(1-x)}) + e^{zxv}e^{z(1-x)} z - (e^z - 1).
\end{align*}
As $h(x^*, z)=0$, we have $\frac{e^{zx^*v}(1-e^{z(1-x^*)})}{x^*-1} = e^z-1$. This implies
\begin{align*}
\frac{\partial h (x^*, z)}{\partial x} >0
& \iff e^{zx^*v} (zv)(1-e^{z(1-x^*)}) + e^{zx^*v}e^{z(1-x^*)} z > (e^z - 1)\\
& \iff zv(x^*-1)+ \frac{z(x^*-1)}{1-e^{z(1-x^*)}}e^{z(1-x^*)} > 1\\
& \iff v > \frac{1}{z(x^*-1)} - \frac{1}{e^{z(x^*-1)}-1}.
\end{align*}
The last inequality always holds because of $v > 1/2$ and Part 1 of \autoref{claim:auxiliary}.
\end{proof}

\begin{claim}
If $x^* < 2$, then $\frac{\partial h (x^*, z)}{\partial z} <0$.
\label{claim:derivative_h_z_negative}
\end{claim}

\begin{proof}
Observe that
$$
\frac{\partial h (x, z)}{\partial z} =e^{zxv} (xv) (1-e^{z(1-x)}) + e^{zxv} e^{z(1-x)}(x-1) - (x-1) e^z.
$$
As $h(x^*, z)=0$, we have $e^{zx^*v} = \frac{(x^*-1)(e^z-1)}{1-e^{z(1-x^*)}}$. This implies
\begin{align*}
\frac{\partial h (x^*, z)}{\partial z} < 0
& \iff (x^*v) + \frac{(x^*-1)e^{z(1-x^*)}}{1-e^{z(1-x^*)}}
 < \frac{e^z}{e^z-1}\\
& \iff (x^*v) + \frac{(x^*-1)}{e^{z(x^*-1)} - 1}
 < 1 + \frac{1}{e^z-1}.
\end{align*}
On the other hand, $h(x^*, z)=0$ is equivalent to
\begin{align*}
zx^*v &= \log\left(\frac{(x^*-1)(e^z-1)}{1-e^{z(1-x^*)}}\right) = \log\left(\frac{z(x^*-1)}{1-e^{z(1-x^*)}}\right) + \log\left(\frac{e^z-1}{z}\right)\\
& = \log\left(\frac{z(x^*-1)}{e^{z(x^*-1)-1}}\right) + z(x^*-1) - \log\left(\frac{z}{e^z-1}\right).
\end{align*}
Therefore,
\begin{align*}
& \frac{\partial h(x^*, z)}{\partial z} < 0 
\iff (zx^*v) + \frac{z(x^*-1)}{e^{z(x^*-1)} - 1}
 < z + \frac{z}{e^z-1}\\
&\iff \log\left(\frac{z(x^*-1)}{e^{z(x^*-1)-1}}\right) + z(x^*-1) + \frac{z(x^*-1)}{e^{z(x^*-1)} - 1}
 < \log\left(\frac{z}{e^z-1}\right)+ z + \frac{z}{e^z-1}.
\end{align*}
The last inequality always holds for $x^*<2$ by Part 2 of \autoref{claim:auxiliary}.
\end{proof}

If $\mu$ is sufficiently small (i.e., a large $z$), then $r^D_{\mu}$ is close to the equilibrium fraction $v > \frac{1}{2}$ for the case of zero cost of information acquisition, and thus $r^D_{\mu} > 1/2$ (i.e., $x^* < 2$). If $x^* < 2$, then $x^*$ increases in $z$ (i.e., decreases in $\mu$) by  \autoref{claim:derivative_h_x_positive}, \autoref{claim:derivative_h_z_negative}, and the Implicit Function Theorem. Hence, as $\mu$ increases (i.e., $z$ decreases), $r^D_{\mu}$ remains to be greater than $1/2$ (i.e., $x^*$ remains less than $2$), and $r^D_{\mu}$ continues to increase in $\mu$.

\subsubsection{Proof for the Boston mechanism}

We assume that $v$ and $\mu$ are chosen such that an equilibrium with fraction $r^{B}_\mu$ exists in the interval $(1/2,1)$. We use a change of variables $x \equiv \frac{1}{r}$, $z \equiv \frac{\lambda_s}{\mu}$, $w \equiv \frac{\lambda_a}{\mu}$, and $y \equiv \frac{\lambda_b}{\mu}$. Then, the equilibrium condition \eqref{eqn:boston_eq_condition_new} becomes
\begin{equation}
e^z = 1 + \frac{e^{zx} -1}{(x-1)e^{\min\{\alpha_1(x), \alpha_2(x)\}} e^{-zvx} +1},
\label{eqn:boston_eq_condition_last_new}
\end{equation}
where $\alpha_1(x) \equiv \frac{w x}{x-1}$ and $\alpha_2(x) \equiv (y+z) x$.

For $i = 1, 2$, let us define:
$$
h_i(x, \mu) \equiv e^{z v x}(1-e^{z(1-x)}) - (x-1)(e^{z}-1)e^{\alpha_i(x) - zx}.
$$
Note that if $x_\mu^{B} \geq \frac{1}{1-\lambda_a}$, then $h_1(x_\mu^{B}, \mu) = 0$. Similarly, if 
$x_\mu^{B} \leq \frac{1}{1-\lambda_a}$, then $h_2(x_\mu^{B}, \mu) = 0$. 

We will now prove that for $i = 1, 2$, if a solution $x_i^{*}$ of $h_i(x ; \mu) = 0$ is less than 2, then the solution strictly decreases as $\mu$ increases. The proof follows a similar approach to the proof for the DA mechanism, where the Implicit Function Theorem is used to show that $r^D_\mu$ increases in $\mu$. To apply the Implicit Function Theorem, we need to make the following claim.

\begin{claim}
For any $x \in (1,2)$ and $y >0$, we have $\log\frac{(x-1)(e^{z}-1)}{1-e^{z(1-x)}} > \frac{zx}{2}$.
\label{claim:boston_aux}
\end{claim}

\begin{proof}
It is sufficient to show that for any $x \in (1,2)$
$$(x-1)(e^y-1) > e^{xy/2} - e^{y(1-(x/2))}.$$
Both sides of the above inequality are equal to $0$ at $x=1$ and equal to $e^y-1$ at $x=2$. The left-hand side is a linear function of $x$, while the right-hand side is a strictly convex function of $x \in (1,2)$. To see the strict convexity, note that the second derivative of the right-hand side with respect to $x$ is given by $(e^{xy/2} - e^{y(1-(x/2))})'' = (y/2)(e^{xy/2} + e^{y(1-(x/2))})' = (y/2)^2 (e^{xy/2} - e^{y(1-(x/2))}) > 0$.
\end{proof}

\begin{claim}
For $i=1, 2$, if $x_i^{*} \in (1,2)$, then $\frac{\partial h_i(x_i^{*}, \mu)}{\partial x} > 0$.
\label{claim:last_heavy_lifting_1}
\end{claim}

\begin{proof}
For each $i=1, 2$,
\begin{align*}
\frac{\partial h_i (x, \mu)}{\partial x} = &e^{z v x} zv(1-e^{z(1-x)}) + e^{zvx}e^{z(1-x)} z \\
& - (e^{z} - 1) e^{\alpha_i-zx} - (x-1)(e^{z} -1)e^{\alpha_i-zx} (\alpha'_i -  z),
\end{align*}
where $\alpha'_1  \equiv \frac{d\alpha_1}{dx}= -\frac{w}{(x-1)^2}$ and $\alpha'_2 \equiv \frac{d\alpha_2}{dx} = y+z$.

Since $h_i(x_i^*, \mu) = 0$, we have $e^{zv x_i^*}(1-e^{z(1-x_i^*)}) = (x_i^*-1)(e^{z} - 1)e^{\alpha_i- zx_i^*}$, by which and $z$ we divide the above derivative and obtain:
\begin{align}
\frac{\partial h_i (x_i^*, \mu)}{\partial x} > 0 & \iff v + \frac{1}{1-e^{z(1-x_i^*)}}e^{z(1-x_i^*)} > \frac{1}{z(x_i^*-1)} + \left(\frac{\alpha'_i}{z} - 1 \right)\notag\\
& \iff v - \left(\frac{\alpha'_i}{z} - 1\right) > \frac{1}{z(x_i^*-1)} - \frac{1}{e^{z(x_i^*-1)}-1}.\label{eqn:last_heavy_lifting}
\end{align}
The right-hand side of \eqref{eqn:last_heavy_lifting} is less than $\frac{1}{2}$ (Part 1 of \autoref{claim:auxiliary}). On the other hand, the left-hand side of \eqref{eqn:last_heavy_lifting} is greater than $\frac{1}{2}$ because, 
\begin{itemize}
\item (for $i = 1$) $v- \left(\frac{\alpha'_1}{z} -1\right) > v + 1 > \frac{1}{2}$, and 
\item (for $i=2$) $h_2(x^*_2, \mu) = 0$ implies $zv x^*_2 = \log \frac{(x^*_2-1)(e^{z}-1)}{1-e^{z(1-x^*_2)}} + (\alpha_2 - zx^*_2)$, and $\alpha_2 =\alpha'_2 x^*_2$. Thus
\begin{align*}
v - \left(\frac{\alpha'_2}{z} -1\right)  = v +1 - \frac{\alpha_2}{z x^*_2} > \frac{1}{2} & \iff \log\frac{(x^*_2-1)(e^{z}-1)}{1-e^{z(1-x^*_2)}} > \frac{zx^*_2}{2},
\end{align*}
and the last inequality holds by \autoref{claim:boston_aux}.
\end{itemize}
Therefore, we conclude that $\frac{\partial h_i (x^*_i, \mu)}{\partial x} > 0$.
\end{proof}

\begin{claim}
For $i=1, 2$, if $x_i^{*} \in (1,2)$, then $\frac{\partial h_i(x_i^{*}, \mu)}{\partial \mu} > 0$.
\label{claim:last_heavy_lifting_2}
\end{claim}

\begin{proof}
For each $i=1, 2$, 
\begin{align*}
\frac{\partial h_i(x, \mu)}{\partial \mu} = & z' \left(e^{zvx} v x (1-e^{z(1-x)}) - e^{zvx}e^{z(1-x)}(1-x) -  (x-1) e^{z}  e^{\alpha_i - zx} \right) \\
& - (x-1)(e^{z}-1)e^{\alpha_i -zx} (\alpha'_i - z' x),
\end{align*}
where $z' = -\frac{\lambda_s}{\mu^2}$, $w' = -\frac{\lambda_a}{\mu^2}$, $y' = -\frac{\lambda_b}{\mu^2}$, $\alpha'_1 = \frac{w' x}{x-1} < 0$, and $\alpha'_2 = (y' + z') x < 0$.

Since $h_i(x^*_i, \mu) = 0$, we have $e^{zv x^*_i}(1-e^{z(1-x^*_i)}) = (x^*_i-1)(e^{z} - 1)e^{\alpha_i-zx^*_i}$, by which and $z'$ we divide the above derivative and obtain:
\begin{align*}
\frac{\partial h_i(x^*_i, \mu)}{\partial \mu} < 0& \iff v x^*_i - \frac{e^{z(1-x^*_i)}(1-x^*_i)}{1-e^{z(1-x^*_i)}} < \frac{e^{z}}{e^{z}-1} + \left(\frac{\alpha'_i}{z'} - x^*_i\right)\\
& \iff zv x^*_i + \frac{z (x^*_i-1)}{e^{z(x^*_i-1)}-1} < z + \frac{z}{e^{z} -1} + (\alpha_i - zx^*_i),
\end{align*}
where the last step follows from $\frac{\alpha'_i}{z'} = \frac{\alpha_i}{z}$ for $i=1,2$.

On the other hand, $h_i(x^*_i, \mu)=0$ implies that
\begin{align*}
& z v x^*_i = \log \left(\frac{(x^*_i-1)(e^{z}-1)e^{\alpha_i - zx^*_i}}{1-e^{z(1-x^*_i)}}\right)\\
& = \log \left(\frac{z(x^*_i-1)}{e^{z(x^*_i-1)}-1}\right) + z(x^*_i-1) - \log \left(\frac{z}{e^{z}-1}\right) + (\alpha_i -zx^*_i).
\end{align*}
Therefore, $\frac{\partial h_i(x^*_i, \mu)}{\partial \mu} < 0$ if and only if 
$$\log \left(\frac{z(x^*_i-1)}{e^{z(x^*_i-1)}-1}\right) + z(x^*_i-1) + \frac{z(x^*_i-1)}{e^{z(x^*_i-1)}-1} < \log \left(\frac{z}{e^{z}-1}\right) + z + \frac{z}{e^{z}-1},
$$
which always holds because of $x_i^{*}<2$ and Part 2 of \autoref{claim:auxiliary}.
\end{proof}

For each $i=1, 2$, if the solution $x^*_i$ of $h_i(x, \mu) = 0$ is less than $2$, then it will decreases as $\mu$ increases due to \autoref{claim:last_heavy_lifting_1}, \autoref{claim:last_heavy_lifting_2}, and the Implicit Function Theorem. As $x^B_\mu = x^*_i$ for either $i=1$ or $i=2$, this implies that $x^B_\mu < 2$ also decreases as $\mu$ increases. Consequently, when $r^B_{\mu} (= 1 / x^B_\mu)$ becomes greater than $1/2$, it will increase as $\mu$ increases.

\bigskip

Next, we will prove that $v \leq \frac{1}{2} + \frac{\lambda_b}{\lambda_s}$ implies $r^B_\mu \leq \max\{\frac{1}{2}, 1-\lambda_a\}$.

Suppose that $r^B_\mu > 1-\lambda_a$. Then, by the equilibrium condition \eqref{eqn:boston_eq_condition_new}, $r^B_\mu$ is the unique solution of 
$$e^z = 1 + \frac{e^{\frac{z}{r}} -1}{\frac{1-r}{r}e^{\frac{y+z}{r}} e^{-\frac{zv}{r}} +1}\equiv \tilde{h}(r).$$
In a previous proof (\autoref{claim:single_dipped_boston_rhs}), we showed that $\tilde{h}(r)$ is single-crossing $e^z$ from above to below as $r$ increases in $[\hat{r}, 1)$ (\autoref{claim:single_dipped_boston_rhs}). Moreover, \begin{align*}
e^z \geq \tilde{h}(1/2) \iff & e^{2(y+z)} e^{-2zv} +1 \geq \frac{e^{2z} -1}{e^z -1} = e^z + 1\\
\iff & 2(y+z-zv) > z \iff v \leq \frac{1}{2} + \frac{y}{z} = \frac{1}{2} + \frac{\lambda_b}{\lambda_s}.
\end{align*}
Therefore, we conclude that $r^B_{\mu} \leq 1/2$.

\bibliographystyle{ecta}
\bibliography{sab_school}
\end{document}